\RequirePackage[2021-06-01]{latexrelease}
\documentclass{aa}
\usepackage{tablefootnote}
\usepackage{graphicx, color}
\usepackage[space]{grffile}
\usepackage{latexsym}
\usepackage{float}
\usepackage{booktabs}
\usepackage{amsfonts}
\usepackage{amsmath}
\usepackage{amssymb}
\usepackage{url}
\usepackage{multirow}
\usepackage{xcolor}
\usepackage[normalem]{ulem}
\usepackage{txfonts}

\title{Cosmology from LOFAR Two-metre Sky Survey Data Release 2: Counts-in-Cells Statistics}
   
\author{Morteza Pashapour-Ahmadabadi$^{1}$\thanks{E-mail: morteza.pasha@physik.uni-bielefeld.de},
Lukas B\"ohme$^{1}$, 
Thilo M.~Siewert$^{1,2}$,
Dominik J.~Schwarz$^{1}$, 
Catherine L.~Hale$^{3,4}$
Caroline Heneka$^{5}$,
Prabhakar Tiwari$^{6}$, 
Jinglan Zheng$^{1}$}

\institute{$^{1}$ Fakult\"at f\"ur Physik, Universit\"at Bielefeld, Postfach 100131, 33501 Bielefeld, Germany \\$^{2}$ Evangelisches Klinikum Bethel gGmbH, Kantensiek 11, 33615 Bielefeld, Germany \\
$^{3}$ Astrophysics, Department of Physics, University of Oxford, Denys Wilkinson Building, Keble Road, Oxford, OX1 3RH, UK \\
$^{4}$  Institute for Astronomy, University of Edinburgh Royal Observatory, Blackford Hill, Edinburgh, EH9 3HJ, UK \\
$^{5}$ Institut f\"ur Theoretische Physik, Universit\"at Heidelberg, Philosophenweg 16, 69120 Heidelberg, Germany \\
$^{6}$ Department of Physics, Guangdong Technion - Israel Institute of Technology, Shantou, Guangdong 515063, P.R. China
}

\authorrunning{Pashapour-Ahmadabadi et al.}

\date{Received XXX ; accepted XXX}
    
\begin{document} 

  \abstract
   {The second data release of the LOFAR Two-Metre Sky Survey (LoTSS-DR2) extends the first data release in terms of sky coverage and source density.  It provides the largest radio source catalogue to date, including 4.4 million sources and covering 5\,635~square degrees of the sky. Therefore, it provides an excellent opportunity for studies of the large-scale structure of the Universe.
   }
   {We investigate the
   statistical distribution of source counts-in-cells and we test a computationally cheap method based on the counts-in-cells to estimate the two-point correlation function.}
   {We study and compare three stochastic models for the counts-in-cells which 
   result in a Poisson distribution, a compound Poisson distribution, and a negative binomial distribution. By analysing the variance of counts-in-cells for various cell sizes, we fit the reduced normalised variance to a single power-law model representing the angular two-point correlation function.}
   {Our analysis confirms that radio sources are not Poisson distributed, which is most likely due to multiple physical components of radio sources. 
   Employing instead a Cox process, we show that there is strong evidence in favour of the negative binomial distribution above a flux density threshold of 2 mJy. 
   Additionally, the mean number of radio components derived from the negative binomial distribution is in good agreement with corresponding estimates based on the value-added catalogue of LoTSS-DR2.
   The scaling of the counts-in-cells normalised variance with cell size is in 
   good agreement with a power-law model for the angular two-point correlation.
   At a flux density threshold of 2 mJy and a signal-to-noise ratio of 7.5 for individual radio sources, we find that for a range of angular scales large enough to not be affected by the multi-component nature of radio sources, the value of the exponent of the power law ranges from $-0.8$ to $-1.05$. This closely aligns with findings from previous optical, infrared, and radio surveys of the large scale structure.
   }
   {The multi-component nature of LoTSS radio sources is essential to understand the observed counts-in-cells statistics. The scaling of the counts-in-cells statistics with cell size provides a computationally efficient method to estimate the two-point correlation properties, offering a valuable tool for future large-scale structure studies.
   }

   \keywords{
   Cosmology: observations,
   large-scale structure of Universe,  Methods: data analysis
   }

   \maketitle
%

\section{Introduction}

The observation of cosmic large-scale structures, like the angular and spatial distribution of galaxies, allow us to relate their statistical properties to models of their formation. The counts-in-cells statistics -- the most primitive statistics one can think of -- holds valuable information about galaxy clustering (\citealp{NeymanShane1953}; \citealp{Peebles1980}; \citealp{saslaw2000distribution}; \citealp{Bernardeau2002}). Despite its significance, it received less attention compared to the more widely studied two-point correlation function (\citealt{Peebles1980}; \citealt{Wang2013}), which for Gaussian random fields contains all non-trivial information (\citealt{Peebles1980}; \citealt{saslaw2000distribution}). Different distribution functions were proposed to model the counts-in-cells distribution of galaxies
(\citealt{NeymanShane1953};
\citealt{ShethSaslaw1994};
\citealt{SaslawYang2011}; \citealt{Hurtago-Gil2017}) and the distribution of matter (\citealt{Anatoly2018}; \citealt{Di2020}).
In this work, we measure the counts-in-cells statistics of radio continuum sources to investigate their statistical distribution and angular clustering, thereby providing insights into the large-scale properties of the cosmic web.

Extra-galactic radio sources, being unaffected by dust extinction and detectable across large cosmological distances, can provide an unbiased sample for probing larger volumes compared to those probed by optical surveys \citep{Zotti2009}. 
Above  a frequency-dependent flux density threshold of about 2 mJy at 150 MHz, the population of radio sources is dominated by Active Galactic Nuclei (AGNs, \citealp{Huynh2008}; \citealp{Rawlings2015}: \citealp{Smolvc2017b}; \citealp{Algera2020}; \citealp{DeepFields_Best2023}). 
With respect to the radio continuum, AGNs fall into two main groups: those with strong radio emissions (radio-loud) and those with faint or absent radio emissions (radio-quiet). 
According to \citet{DeepFields_Best2023}, above about 2 mJy at 150 MHz, a significant fraction of sources are radio-loud AGNs. At sub-mJy flux densities, radio-quiet AGNs are an important class, but accounting for only less than 10 percent of all sources.
Below 1 mJy, star-forming galaxies (SFGs) become the dominant population, comprising 90 percent of sources at flux densities of about 0.1 mJy, but make up less than 20 percent of all sources above a flux density of 2 mJy \citep{DeepFields_Best2023}. 
Through surveys covering wide areas of sky, radio galaxies and quasars can be detected over significant cosmological distances (up to $z \sim 7$, see e.g. \citealt{DeBreuck2010}, \citealt{Singh2014}, \citealt{Saxena2019},  \citealt{Sotnikova2021}, \citealt{duncan2021}, \citealt{DeepFields_Best2023}). 
This offers a reliable way to study the distribution and clustering of galaxies. Galaxy clustering quantifies the excess probability of finding galaxies at a given spatial separation (spatial clustering) or angular separation (angular clustering) compared to a random distribution (\citealp{Totsuji1969}, \citealp{peebles1968}, \citealp{Peacock1991}, \citealp{Cress1996}, \citealp{blake2002}, \citealp{Wang2013}).
Because most radio galaxies have very faint optical counterparts, measuring their redshift is difficult. Therefore, many studies have been conducted on the angular clustering of radio continuum sources across the radio spectrum (see \citealt{SlenderPeebles1981} for the fourth Cambridge survey (4C), \citealt{Magliocchetti1999} for the Faint Images of the Radio Sky at Twenty-Centimetres (FIRST) survey, \citealt{WENESS_GB6} for the Westerbork Northern Sky Survey (WENSS) and  the Green Bank 6-cm (GB6) survey, \citealt{NVSS2002} for the NRAO VLA Sky Survey (NVSS), \citealt{TGSS_Clustering} for the TIFR GMRT Sky Survey (TGSS), \citealt{Siewert2019} and \citealt{Hale2023} for the LOFAR Two-metre Sky Survey (LoTSS) first and second data release (DR1, DR2), respectively).

Counts-in-cells statistics have received less attention but have also been studied with different radio surveys (\citealp{Jauncey1975}; \citealp{Magliocchetti1999} for the FIRST survey; \citealp{Siewert2019} for LoTSS-DR1). \citet{Siewert2019} have shown that the distribution of radio sources is non-Poissonian and that a compound Poisson distribution provides a better description. In this work, we probe two theoretical model distributions (compound Poisson and negative binomial distributions) to determine the best fit for the distribution of these sources. Using moments of the counts-in-cells -- such as the mean and variance -- we study the cosmic large scale structure as traced by continuum radio sources from the second data release (DR2) of LoTSS \citep{Shimwell2022} by increasing the area of the survey by approximately a factor of 10 compared to LoTSS-DR1 \citep{LoTSSDR1A}.

The Low-Frequency Array (LOFAR; \citealt{LOFAR2013}) is a large radio interferometer designed to observe the entire northern sky at frequencies between 10 and 240 MHz. LOFAR consists of core, remote, and international stations, with the core and remote stations located in the Netherlands and the international stations spread across Europe. LOFAR observes the radio sky using two types of antennas: LBA (Low Band Antennas: 10–90 MHz) and HBA (High Band Antennas: 110–240 MHz).  
The LOFAR Two-metre Sky Survey (LoTSS; \citealt{LoTSS2017}; \citealt{LoTSSDR1A}; \citealt{Shimwell2022}), covering frequencies between 120 and 168 MHz, makes use of the HBAs. The central observing frequency of LoTSS is 144 MHz, which corresponds to a wavelength of about 2 metres. Thanks to its sensitivity, resolution and large sky coverage (detailed further in Sect. \ref{sec: data_quality}), LoTSS provides an excellent opportunity to investigate the statistical properties of the radio sky.

In Sect.~\ref{sec: countsincells_theory}, we provide an overview of the theoretical frameworks that encompass the various distributions employed in this analysis. We assess the data quality of LoTSS-DR2 in Sect.~\ref{sec: data_quality} and the mock random catalogue in Sect.~\ref{sec:mock}. Sect.~\ref{sec:masking_strategies} discusses the masking strategies used, while Sect.~\ref{sec:DifferentialCounts} presents the differential source counts and completeness. In Sect.~\ref{sec:results}, we outline the results of the one-point correlation functions of LoTSS-DR2 and explore the scaling properties of the cell sizes to find the two-point correlation function. Finally, in Sect.~\ref{sec: conclution}, we offer conclusions and engage in further discussions. Five appendices cover several aspects of masking (App.~\ref{sec:Masks} and \ref{sec:threecells}),  
comparison to LoTSS-DR1 in App.~\ref{app:DR1} and present further technical details (App.~\ref{sec:Geometry-based_masks_results} and \ref{sec:cgamma}).

\begin{figure*}
\centering
\includegraphics[width=\linewidth]{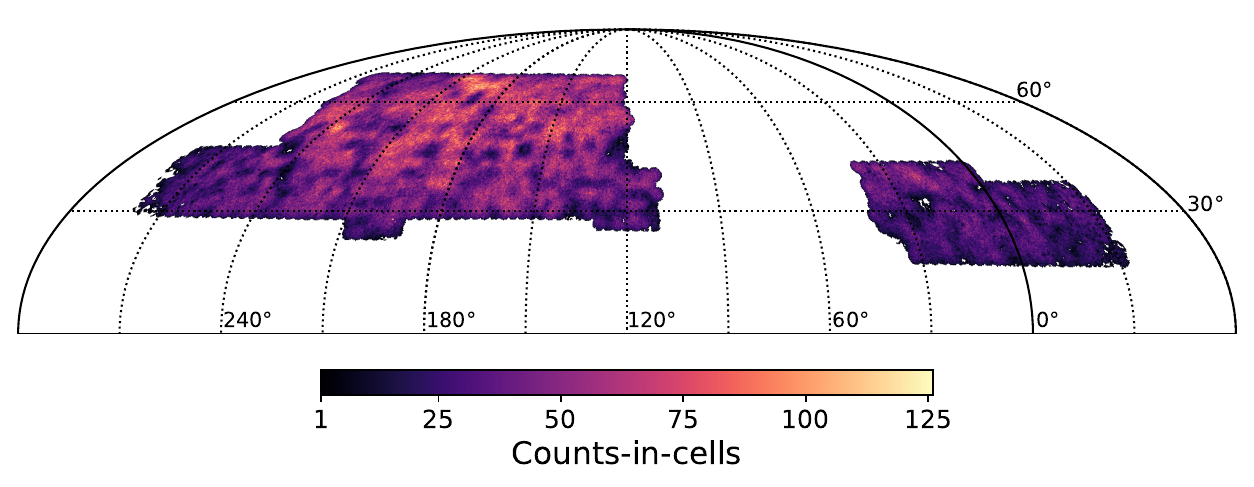}
\caption{Counts-in-cells of the LoTSS-DR2 radio source catalogue in Mollweide view and equatorial coordinates without a flux density cut. The counts-in-cells are based on {\sc HEALPix} with a resolution of 13.74  square arcmin.}
\label{fig:fullDR2}
\end{figure*}

\section{Counts-in-cells}
\label{sec: countsincells_theory}

The counts-in-cells statistics is a simple method to study the spatial distribution of sources within a survey region (see  e.g. \citealp{NeymanShane1953},  \citealp{Peebles1980}, \citealp{Magliocchetti1999},   \citealp{saslaw2000distribution},  \citealp{Bernardeau2002}). This approach involves dividing the survey area into smaller, equally sized cells and counting the number of sources within each cell (counts-in-cells). These cells are equivalent to bins in a histogram. We use counts-in-cells statistics to explore the underlying distribution of the sources in LoTSS-DR2. The distribution can provide insight into the fundamental cosmological physics and thermodynamic processes of the Universe (\citealp{SaslawYang2011}; \citealp{Hurtado-Gil2017}). 

The cosmological principle states that the statistical distribution of matter is isotropic (the same in all directions) and homogeneous (uniform) when observed on scales larger than hundreds of megaparsecs. If radio sources would be statistically independent tracers of the cosmic matter distribution, we would expect them to be distributed isotropically. Disregarding any imperfections of the observations and systematic biases in the survey pipelines, and assuming that all sources are point-like (i.e., their angular size is smaller than the angular resolution of the telescope), as well as statistically independent, the counts-in-cells statistics should follow a Poisson distribution (\citealp{Poisson2014}). 

However, that ignores the fact that many astrophysical objects show up as multi-component sources, e.g. the core and lobes of AGNs, or resolved large spirals with many individual components. Examples are provided in App.~\ref{sec:threecells}, including the resolved LoTSS image of M101 (Pinwheel) galaxy, which the source-finding algorithm fails to identify as a single object. While large prominent objects like M101 can easily be manually corrected to contribute as a single source to the source counts, this correction becomes far more challenging for fainter radio sources. Additionally, there are non-physical associations, like image artefacts, or associations at larger scales, like several members of a galaxy cluster or groups of galaxies, as well as cosmological structure formation causing correlations on all scales. All of those lead to deviations from a Poisson distribution. Previously, \citet{Siewert2019} showed that deviations from a Poisson distribution are found with high statistical significance. This work aims to extend that analysis by increasing the survey area by approximately a factor of 10.

\subsection{Statistical moments}\label{sec:statmoments}

We briefly recall the sample statistics which have been used in \citet{Siewert2019} to describe the distribution of the source counts $k_i$ per cell $i$ (with $N_\mathrm{cell}$ disjoint cells of identical solid angle). The statistical moments, which are numerical measures describing the shape and characteristics of a probability distribution, are estimated as
\begin{equation} 
\label{eq:moments}
\mu_j \equiv \frac{1}{N_\mathrm{cell}}\sum_{i=1}^{N_\mathrm{cell}}k_i^j, 
\end{equation} 
with the sample mean $\mu \equiv \mu_1$. Then the 
central moments, which describe the shape of the distribution by taking deviations from the mean, are
\begin{equation} 
\label{eq:centralmoments}
m_j \equiv \frac{1}{N_\mathrm{cell}}\sum_{i=1}^{N_\mathrm{cell}}(k_i-\mu)^j, 
\end{equation} 
with the variance $\sigma^2 \equiv m_2$,
and the coefficients of skewness and excess kurtosis \citep{ZwillingerKokoska2000}
\begin{equation} 
\label{eq:3}
g_1 \equiv \frac{m_3}{m_2^{3/2}}, \quad
g_2 - 3 \equiv \frac{m_4}{m_2^2}-3,
\end{equation}
respectively. Skewness measures the degree of asymmetry in a distribution, while kurtosis measures the tailedness of a distribution and is therefore very sensitive to outliers.
We also define the clustering parameter \citep{Peebles1980},
\begin{equation}
\label{eq:4}
    n_c \equiv \frac{\sigma^2}{\mu},
\end{equation}
which is unity for a Poisson distributed quantity and where values of $n_c$ exceeding unity indicate clustering. For a Poisson distribution we expect 
\begin{equation}
g^\mathrm{P}_1 =\mu^{-1/2}, \quad g^\mathrm{P}_2 - 3 = \mu^{-1}.
\end{equation} \label{eq:5}

\subsection{Modelling the counts-in-cells distribution}\label{sec:cic_dist}

Deviations from a Poisson distribution are expected due to various factors. Deviations which are uncorrelated at large angular scales, can stem from either local physics (like in the case of AGNs or nearby spirals), but also imaging artefacts -- such as sidelobes in interferometric data or noise spikes. These effects can be described by what is called a Cox process \citep{Cox1955}, which models a random process within an underlying Poisson random process.

Let $N_i$ be a random variable that denotes the counts of radio sources in a single cell $i$ (measured by means of a radio source catalogue) and 
$O_i$ being the Poisson distributed number of physical objects (i.e.\ an AGN or a SFG) in that cell, then
\begin{equation}
    N_i \sim \sum_{j=1}^{O_i} C_{ji}, \quad O_i \sim \mathrm {Poisson}(\lambda),
\end{equation}
where $C_{ji}$ denotes another random variable that counts the number of components that are associated with a physical object $j$ in cell $i$. Finally, $N = \sum N_i$ denotes the total number of radio sources in the source catalogue.
 
The resulting distribution for $N_i$ (and $N$) depends on the detailed assumptions that we make on the discrete distribution of $C_{ji}$. The associations captured by the component counts $C_{ji}$ can have many different reasons; they can include imaging artefacts, typically associated to bright sources, they could be the lobes correlated with the core of an AGN, or it could happen that a large spiral is broken up in several components by the source-finding algorithm. From the LoTSS-DR1 value-added catalog (\citealt{williams2019}), visual classification in the citizen science project LOFAR Galaxy Zoo has shown that approximately 13,200 sources, or about 4\% of the total, are associated together. This indicates that the distribution of component counts depends on a combination of astrophysical correlations, survey properties, and the specifics of the source-finding algorithm. Given the high level of complexity of these factors, we do not attempt to derive the distribution of $C_{ji}$ from first principles; instead, we base reasonable probability distributions for them on an educated guess. 

If $C_{ji} = 1$ for all objects, $N_i$ and $N$ are Poisson-distributed numbers. \citet{Siewert2019} argued that the counts of components is a Poisson process itself with $C_{ji} \sim$ Poisson($\kappa$), where $\kappa$ denotes the intensity of the process, i.e.\ the mean number of component counts per physical object. This ansatz does allow for zero components, which would mean that we make the assumption that several physical objects remain undetected in the survey -- either due to being below the detection threshold or due to observational limitations. The resulting distribution of $N_i$ can be easily obtained starting from the generating function of a Poisson (P) distribution \citep{johnson2005univariate}, 
\begin{equation}
    G_\mathrm{P}(z) = \exp[\lambda (z - 1)],
    \label{eq:GFP}
\end{equation}
where $\lambda > 0$ denotes the intensity of the Poisson process and $z$ denotes the random variable. Then the generating function of the compound Poisson (CP) distribution becomes
\begin{equation}
    G_\mathrm{CP}(z) = \exp[\lambda (\exp[\kappa (z -1)] - 1)],
    \label{eq:GFCP}
\end{equation}
and we can calculate its mean and variance \citep{johnson2005univariate}, 
\begin{eqnarray}
    \mu_\mathrm{CP} &=& G_\mathrm{CP}'(1) = 
    \lambda \kappa, \\
    \sigma^2_\mathrm{CP} &=& G_\mathrm{CP}''(1) + G_\mathrm{CP}'(1) - [G_\mathrm{CP}'(1)]^2 = \lambda \kappa (1+ \kappa) 
\end{eqnarray}

Consequently, the clustering parameter becomes $n_c^\mathrm{CP} = 1 + \kappa > 1$.
Similarly, we can estimate higher central moments and find 
\begin{eqnarray}
    g_1^\mathrm{CP} &=& \frac{1 + 3 \kappa + \kappa^2}{( \lambda \kappa)^{1/2}(1+ \kappa)^{3/2}} = \frac{n_c^2 + n_c -1}{\mu^{1/2}\, n_c^{3/2}}, \\ g_2^\mathrm{CP} - 3 &=& \frac{1 + 7 \kappa + 6 \kappa^2 + \kappa^3}{\lambda \kappa (1 + \kappa)^2} = \frac{n_c^3 + 3 n_c^2 - 2 n_c - 1}{\mu\, n_c^2}.
\end{eqnarray}

The assumption that the counts of components may turn out to be zero might be in contradiction with the assumption that a survey is complete above a certain flux density. 
Thus, we also make use of a logarithmic distribution of $C_{ji}$, for which the number of components is at least $1$ and is discrete.\footnote{\citet{Fisher43} introduced the logarithmic distribution studying the relation of the number of species and the number of individuals in a random animal population. From the statistical perspective our problem at hand is similar -- with individuals (radio components) that belong to different species (physical objects like AGNs or SFGs).} 
We now show that for 
$C_{ji} \sim$  Logarithmic(p), with 
$0 < p < 1$ denoting the parameter of the distribution, the resulting Cox process produces a negative binomial distribution. This outcome is significant because the negative binomial distribution is well-suited for modelling count data with overdispersion, where variance exceeds the mean, and provides a more realistic representation of the observed radio source counts.
The generating function for the logarithmic (L) distribution \citep{johnson2005univariate} reads
\begin{equation}
    G_\mathrm{L}(z) = \frac{\ln (1-pz)}{\ln (1 - p)}, 
\end{equation}
and the generating function for a Cox process with logarithmic distribution becomes 
\begin{equation}
    \exp[\lambda (\frac{\ln (1-p z)}{\ln (1 - p)} - 1)].
\end{equation}
Introducing the new variable $r \equiv - \lambda/\ln(1-p) > 0$ we find that this becomes the generating function of a negative binomial (NB) distribution \citep{johnson2005univariate}
\begin{equation} 
    G_\mathrm{NB}(z) = \left( \frac{1- p}{1-p z}\right)^r, 
    \label{eq:GFNB}
\end{equation}
from which we obtain the mean and variance, 
\begin{eqnarray}
    \mu_\mathrm{NB} &=& \frac{r p}{1-p}, \\
    \sigma^2_\mathrm{NB} &=& \frac{r p}{(1-p)^2}.
\end{eqnarray}
Therefore, the clustering parameter reads $n_c = 1/(1-p) > 1$ and we find
\begin{eqnarray}
    g_1^\mathrm{NB} &=& \frac{1 +p}{(r p)^{1/2}} = \frac{2 n_c - 1}{\mu^{1/2}\, n_c^{1/2}}, \\ g_2^\mathrm{NB} - 3 &=& \frac{1 + 4 p + p^2}{r p} = \frac{6 n_c^2 - 6 n_c + 1}{\mu\, n_c}.
\end{eqnarray}
In the limit $n_c \to 1$, both the negative binomial and compound Poisson distributions converge to the Poisson distribution, meaning their higher moments (skewness, kurtosis and beyond) also agree with each other.
For $n_c > 1$ and identical mean $\mu$, the negative binomial distribution has larger skewness and excess kurtosis.

The mean number of components per physical object in a negative binomial distribution is given by the expectation value of the logarithmic distribution,
\begin{equation} 
\label{eq:nb_components}
   \mu_L = \frac{ - p}{(1 - p) \ln(1 - p) }.
\end{equation}

A simple method to estimate the parameters of the distributions in Eq.~(\ref{eq:GFP}), (\ref{eq:GFCP}), and (\ref{eq:GFNB}) is to use the estimates of the first and second moments, $\hat\mu$ (mean) and $\widehat{\sigma^2}$ (variance), which provide the following parameter estimates; $\hat \lambda_\mathrm{P} = \hat \mu$ for a Poisson distribution, 
and for the two Cox processes we find that 
\begin{equation}
\hat \lambda_\mathrm{CP} = \hat r_\mathrm{NB} = \frac{{\hat \mu}^2}{\widehat{\sigma^2} - \hat \mu},
\end{equation}
and 
\begin{equation}
\hat{\kappa}_\mathrm{CP} = \frac{\widehat{\sigma^2}}{\hat \mu}
-1, 
\end{equation}
and 
\begin{equation}
\hat{p}_\mathrm{NB} = 1- \frac{\hat \mu}{\widehat{\sigma^2}},
\end{equation}
respectively, where the hat notation denotes the estimator of the parameter. Thus we see that based on the methods of moments, $\lambda_\mathrm{CP}$ and $r$ have the same numerical value. However, if we instead use the method of least square deviations, we do not expect that to happen in general.

\subsection{Angular correlation function from the reduced normalised variance}
\label{sec:scalingofcellsizestheory}

The variance of counts-in-cells depends on the size of the cells in a specific way, which connects this statistics with the angular two-point correlation function, $w(\vartheta)$, as was originally shown in \citet{Totsuji1969} and further elaborated in \citet{Peebles1975}. In essence, the two-point correlation function $w(\vartheta)$ describes how the probability of finding two sources at an angular separation $\vartheta$ deviates from a purely random (Poisson) distribution. Since the variance of source counts is influenced by clustering, it naturally reflects information from $w(\vartheta)$. In general, the statistical $n$th-moment (see Eq.~\ref{eq:moments}), can be written in terms of $n$-point correlation functions, which describe the degree of clustering at different scales. For the second moment ($n=2$), it is useful to consider the reduced and normalised variance
\begin{equation}\label{eq:23}
     \Psi_2 \equiv \frac{m_2 - \mu}{\mu^2} = \frac{1}{\Omega_c^2} \int \!\! \int \; w (\vartheta_{12}) \,\mathrm{d}\Omega_1 \,\mathrm{d}\Omega_2,
\end{equation}
where $\vartheta_{12}$ is the angular distance between two solid angle elements, $\mathrm{d}\Omega_i$, and $\Omega_c$ denotes the area of the cell. When sources are Poisson distributed, $w(\vartheta) = 0$ and $\Psi_2 $ vanishes.
However, in the presence of clustering, correlations between sources lead to deviations from Poisson statistics, making $\Psi_2 $ a crucial measure of the distribution of structures in the Universe.

At small angular scales, the assumption of a power law for the two-point correlation function can be made,
\begin{equation}
     w(\vartheta_{12}) = A_0 \; \left(\frac{\vartheta_{12}}{\vartheta_0}\right)^{1 - \gamma},
\end{equation}
with amplitude $A_0$ defined at the pivot scale $\vartheta_0$, and clustering exponent $\gamma$, which characterizes how the correlation decays with increasing separation $\vartheta_{12}$. Observational studies suggest a typical value of $\gamma \approx 1.8$ (see e.g. \citealt{Peebles1975}, \citealt{blake2002}, \citealt{Willman2003}, \citealt{Lindsay2014}). A higher $\gamma$ means sources are highly clustered at small scales but become uncorrelated at larger scales.

Evaluating the integral in Eq.~(\ref{eq:23}) over a cell, occupying a solid angle $\Omega_c$ with linear angular scale $\Theta \equiv \sqrt{\Omega_c}$, one finds
\begin{equation}
     \Psi_2(\Theta)
     = A_0 \, C_{\gamma} \left(\frac{\Theta}{\Theta_0}\right)^{1 - \gamma},
\end{equation}
where $C_{\gamma}$ is a coefficient of order unity depending on $\gamma$, 
with $C_1 = 1$. $\Theta_0$ serves as a pivot scale. More details on the numerical evaluation of $C_{\gamma}$ are given in  App.~\ref{sec:cgamma}. A higher $\gamma$ means clustering is stronger at small $\Theta$, leading to higher variance at smaller cell sizes.


\section{Data} 

\subsection{LOFAR Two-metre Sky Survey: DR1 \& DR2}\label{sec: data_quality}

The LOFAR Two-metre Sky Survey will eventually cover the most of the northern sky at 120 -- 168 MHz. The first data release \citep{LoTSSDR1A} included 58 pointings in the region of the HETDEX Spring Field and covered 424 square degrees. The second data release \citep{LoTSS2}, which is used in this study, extends the first data release in terms of sky
coverage and source density. It consists of 841 pointings with a coverage of $5\,635$ square degrees of northern sky. Due to resource limitations and ongoing technical developments, only the core and remote stations are used. With maximum baselines extending up to $\sim 100$ km across the Netherlands, the survey achieves an angular resolution of 6 arcseconds. 

Observations for LoTSS are carried out with the LOFAR-HBA (High Band Antenna), utilising the core and remote stations. The pointing size (given by the station beam) of these observations is $3.8$~deg in diameter at $150$ MHz, while the pointings are typically separated by $2.58$~deg, with six nearest neighbours within $2.8$~deg \citep{LoTSSDR1A}. This significant overlap between the pointings can be used for mosaicing – the process of averaging data from overlapping regions – and thereby avoiding lower quality data at the outer edges. In LoTSS-DR2 \citep{LoTSS2}, 626 and 215 pointings cover the RA-13~h and RA-1~h regions, respectively.
The $4\,396\,228$ detected radio sources of LoTSS-DR2 are shown as a {\sc HEALPix}\footnote{\url{http://healpix.sourceforge.net}} \citep{Healpix} counts-in-cells map with a resolution of $N_\mathrm{Side}=256$ as a Mollweide projection in Fig.~\ref{fig:fullDR2}. This resulting catalogue comes from the combined mosaic of the individual pointings where overlapping areas are averaged and combined.

\citet{Shimwell2022} tested the completeness of the radio source catalogue by injecting point-sources 
and real deconvolution components of observed point-sources into the 841 source-subtracted mosaics in 10 simulations. For point-sources a completeness of $50\%$, $90\%$ and $95\%$ is found at $0.34$~mJy, $0.8$~mJy, $1.1$~mJy. This suggests that the investigation of cosmological questions calls for a flux density threshold well above $1.1$~mJy to ensure a completeness level above $95\%$ in all fields. We explore the completeness considerations in greater detail in Sect. \ref{sec:DifferentialCounts}.

\subsection{Random mock catalogue}
\label{sec:mock}
 
For the comparison of the observations with a Poissonian source distribution, we use the random mock catalogue generated by \citet{Hale2023}. As outlined in Sect. 3.2 of \citet{Hale2023}, the generation process involves several sequential steps. Initially, random sources are generated across the survey field of view, excluding regions  
affected by data reduction failures. Next, the process accounts for position-dependent smearing effects, which affect the detection of sources, necessitating modelling and correction for its dependence on field elevation. The mock catalogue also incorporates incompleteness and measurement errors, accounting for sensitivity variations across the survey area and their impact on source detection completeness and properties like flux density. We use it as a comparison to the distribution of the sources and for the estimation of the two-point correlation function (discussed in Sect. \ref{sec:results}).

\begin{figure}
    \centering
    \includegraphics[width=0.98\linewidth]{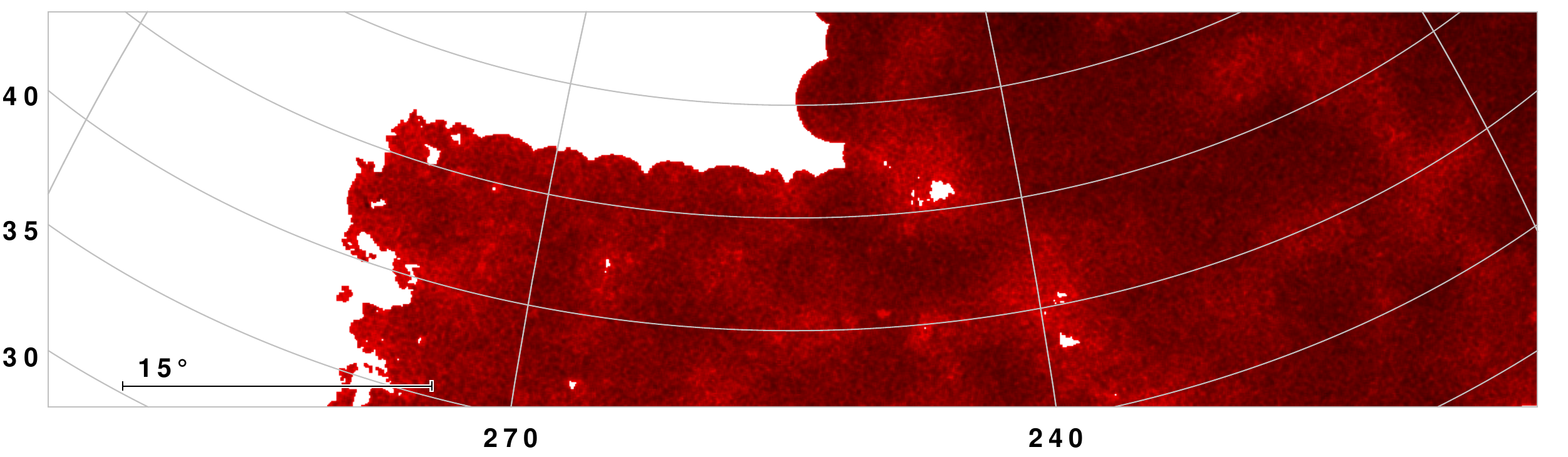}
    \caption{Example of complete pointings in 
    the HETDEX field and incomplete pointings at the boundary of the survey area, and pointings which are not mosaiced with other neighbouring pointings.}
    \label{fig:boundary_poinitings}
\end{figure}

\section{Masking strategies}\label{sec:masking_strategies}

In addition to applying a flux density threshold to ensure completeness (see Sect.~\ref{sec:DifferentialCounts}), we have to exclude regions that are strongly affected by systematics.
Alternative strategies based on the rms noise level percentiles which were employed in \cite{Siewert2019}, and new strategies based on the pointing geometry, tested for DR2, are discussed in App.~\ref{sec:rmsmasks}. 
In the following, we discuss the strategy used to define a mask that excludes the outer regions of the survey, which are affected by systematics, resulting in our default mask.

As discussed in Sect.~3.3.2 and Fig.~9 of \citet{Shimwell2022}, the ratio between the integrated flux density values of LoTSS to NVSS varies systematically as a function of position on the sky. Variations in the flux density scale across an individual pointing are reduced significantly by mosaicing. Through mosaicing the overestimation of flux density in the northern regions of a pointing is somewhat reduced by the underestimation in the southern regions of a neighbouring pointing. The overlapping areas between pointings thus reduce the flux density variation by averaging out these systematic biases across the survey area.

As such, for a mask based on the pointing geometry, it is important that we exclude the outer edges of the survey area, where adjacent pointings have not been mosaiced. These un-mosaiced regions can introduce inconsistencies, such as flux density variations. Therefore, we mask these outer edges of boundary pointings of the survey. Additionally, and for a similar reason, we want to remove those areas where there are a large number of gaps within the image due to facets that failed during the data reduction process (see Fig.~\ref{fig:boundary_poinitings}). 
These incomplete pointings often, though not exclusively, lie towards the outer edges of the observations affecting the amount of masking used. Consequently, we mask both the un-mosaiced boundary areas and the regions with incomplete pointings. This mask is referred as `mask d', which has been adopted from \citet{Hale2023}. 

\begin{figure}
    \centering
    \includegraphics[width=\linewidth]{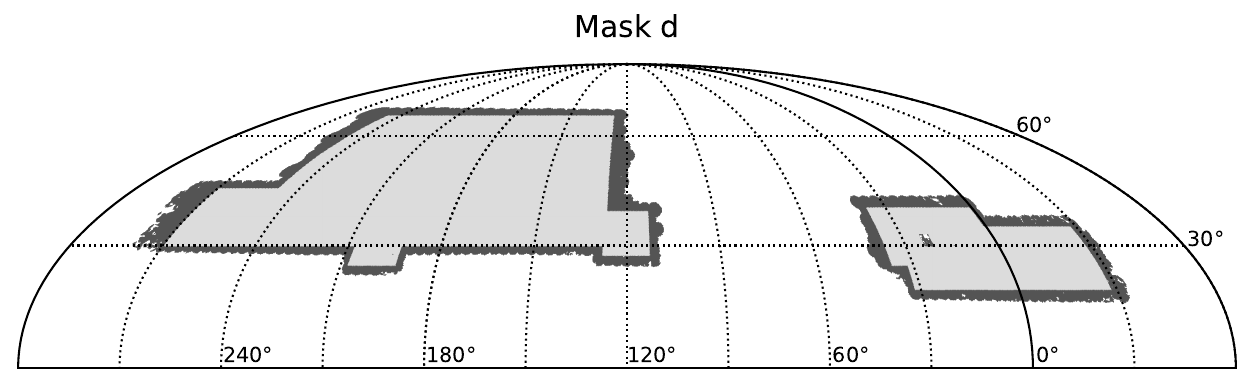}
    
    \caption{For `mask d' the light grey regions remain after masking, while dark grey is the survey area.}
    \label{fig:1017_1219masks}
\end{figure}

This default mask (mask d) is used for all LoTSS-DR2 cosmology analysis and is shown in Fig.~\ref{fig:1017_1219masks}, where we compare it to the sky coverage of the full dataset. The RA and Dec cuts used to determine these regions are given in Table~\ref{tab:regions}. Whilst slightly larger areas could have been included, these cuts are employed to be conservative to ensure that we only analyse regions with reliable data. The RA and Dec cuts are applied to the central position of {\sc HEALPix} pixels, not to the individual sources as done in \cite{Hale2023}. Additionally, we exclude three cells with outstandingly high source density due to large foreground sources split into several single components. For more details see App.~\ref{sec:threecells}. With these cuts applied, we have $\sim$78\% of the total area of LoTSS-DR2 remaining.

\begin{table}
    \centering
    \caption{Definition of 'mask d' used to mask both the data and random catalogues as described in Sect.~\ref{sec:masking_strategies} (following the Table 1 of \citealt{Hale2023}). Regions outside of these boundaries will be masked.}
    \begin{tabular}{ccc}
        \hline \hline
        Region & RA & Dec \\
        & (deg) & (deg)\\\hline 
        1 & [1, 37] & [25, 40] \\
        2 & [1, 32] & [19, 25] \\
        3 & [0, 1] & [19, 35] \\
        4 & [113, 127] & [27.5, 39]\\
        5 & [127, 248] & [30, 67] \\
        6 & [193, 208] & [25, 30] \\
        7 & [248, 270] & [30, 45] \\
        8 & [332, 360] & [19, 35] \\\hline
    \end{tabular}
    \label{tab:regions}
\end{table}

\section{Differential Source Counts} \label{sec:DifferentialCounts}

\begin{figure*}
    \centering
    \includegraphics[width=0.49\linewidth]{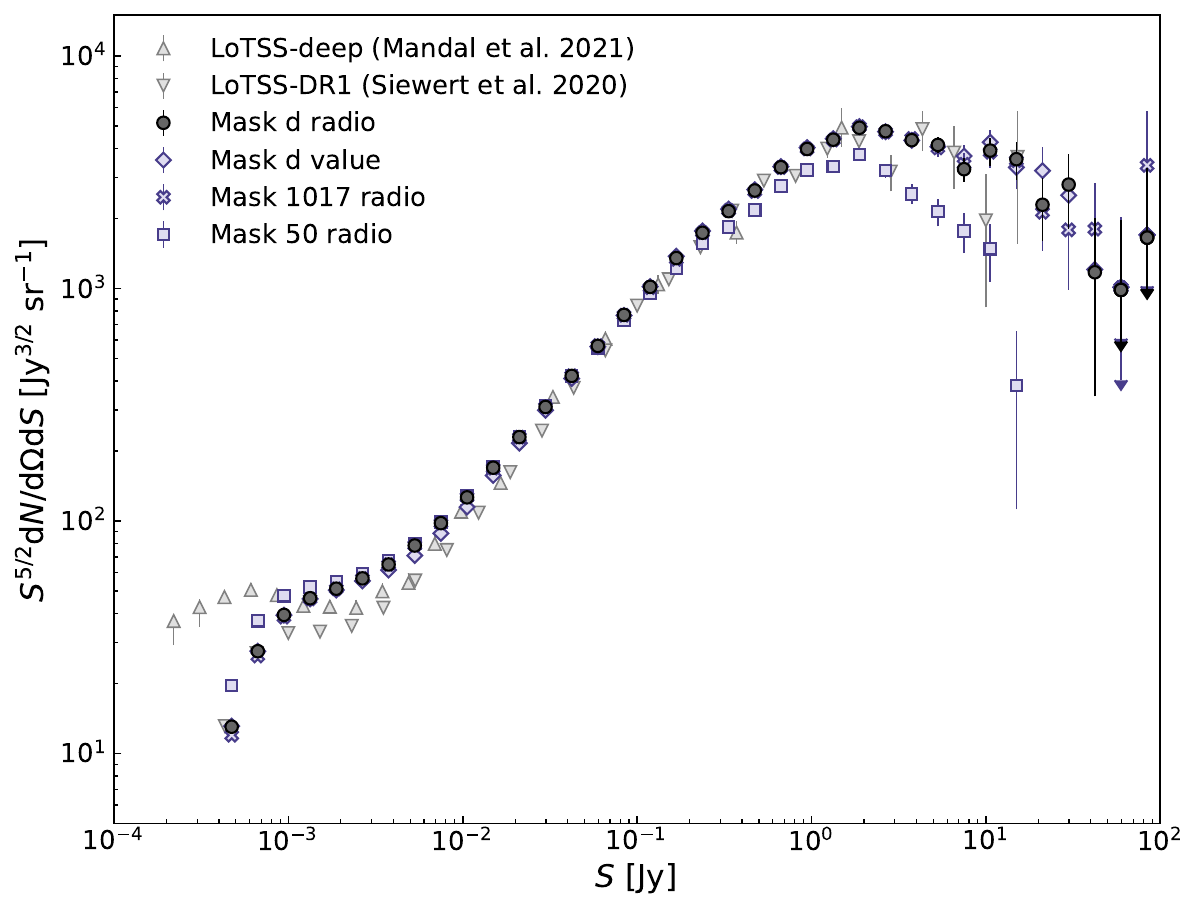}
    \includegraphics[width=0.49\linewidth]{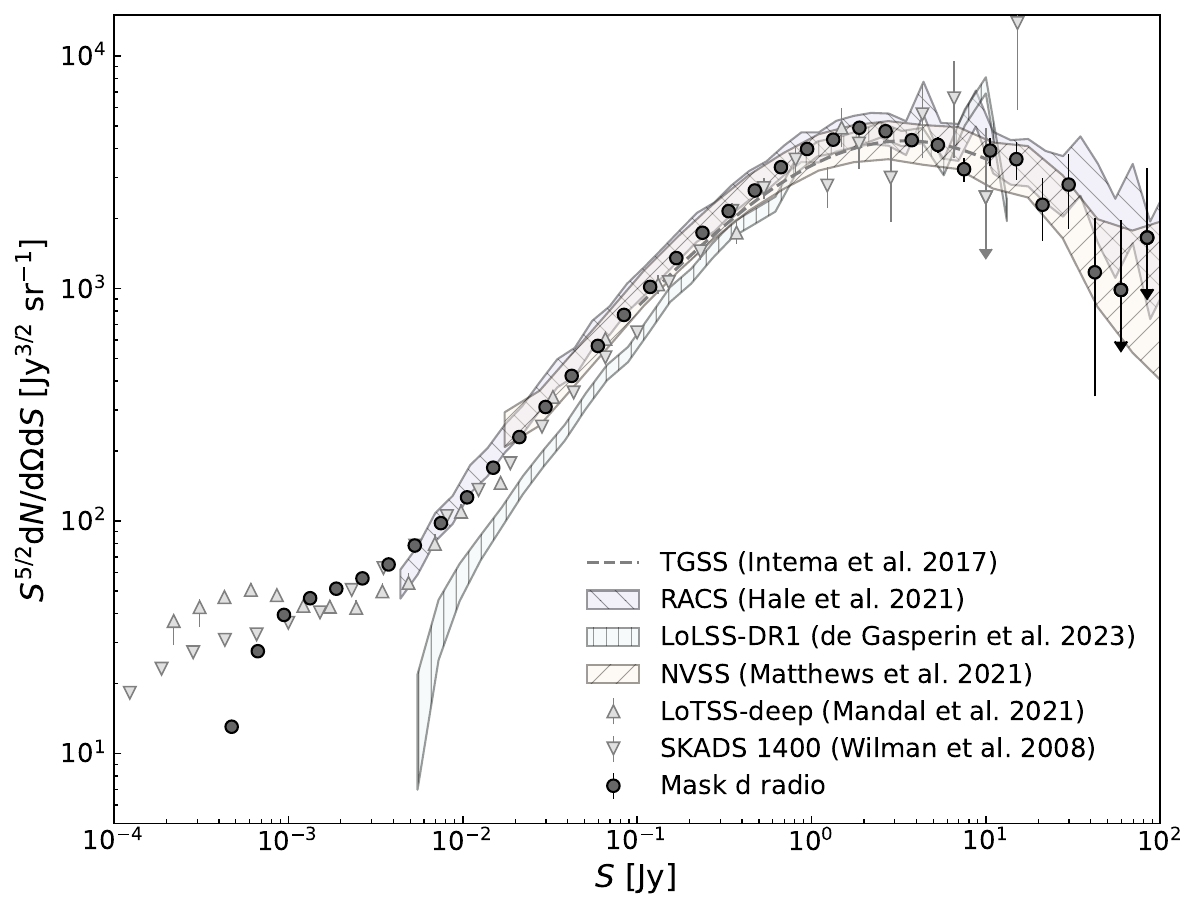}
    \caption{\textit{Left:} Comparison of the differential source counts from mask d, mask 1017 (explained in App.~\ref{sec:pointingGeometry}), and mask 50 (explained in App.~\ref{sec:rmsmasks}) with the results from the LoTSS Deep Fields \citep{mandal2021}, and LoTSS-DR1 \citep{siewert2021}. The ending `radio' refers to the application of the mask to the LoTSS-DR2 radio source catalogue, while the ending `value' refers to the LoTSS-DR2 value-added catalogue. Downward pointing arrows indicate error-bars that extend outside of the plotted y-range. \textit{Right:} Comparison of the differential source counts from `mask d' with the results from the LoTSS Deep Fields, TGSS \citep[150 MHz;][]{intema2017}, RACS \citep[888 MHz;][]{hale2021}, LoLSS-DR1 \citep[54 MHz;][]{deGasperin2021}, NVSS \citep[1.4 GHz;][]{matthews2021}, and SKADS \citep[1400 MHz][]{Wilman2008}.}
    \label{fig:diff_source_counts}
\end{figure*}

We use differential source counts (\citealt{Condon1988}) as a function of flux density to evaluate the completeness of our data and the effectiveness of the applied masks.
In Fig.~\ref{fig:diff_source_counts}, we show the differential number counts using Euclidean normalization, which, in a static, homogeneous, and spatially flat Universe, are expected to remain constant as a function of flux density. In the left panel, we present the differential number counts for `mask d' alongside two more conservative masks, which were derived using different masking strategies (see Sect.~\ref{sec:masking_strategies} and App.~\ref{sec:Masks} for further details on the masks). We also show the `mask d value', which denotes the previously defined `mask d' but applied specifically to the LoTSS-DR2 value-added catalogue \citep{hardcastle2023}. The value-added catalogue combines radio sources, which are observed as several components, into one single source. This process is only complete above 4mJy. As a comparison, we also include the results from the LoTSS Deep Fields \citep{mandal2021} and results from the LoTSS-DR1 value-added catalogue \citep{Siewert2019}.

At low flux densities (on the mJy level) all masks except `mask 50' yield consistent results, which confirms a high degree of completeness for LoTSS-DR2 at $S > 2$mJy. The noise-based mask `mask 50', which keeps the first 50 percent of all cells ranked by increasing averaged cell noise (more in App. \ref{sec:rmsmasks}) retains more sources at a low flux density but removes sources at higher flux densities, likely due to higher rms noise around bright sources, and therefore explains the difference at the lower and higher flux densities.
The value-added catalogue shows a lower source count in the flux density range between 4 and 20 mJy, which is due to the value-adding process (see the left panel of Fig.~\ref{fig:diff_source_counts}).

In addition to the completeness analysis discussed above, we can confidently rely on the source counts of mask d for flux density levels above 2 mJy (see the right panel of Fig.~\ref{fig:diff_source_counts}). Based on the differential source counts, we find that the more sophisticated masking strategy (for more details, see App.~\ref{sec:Masks}) does not offer any advantage over the simpler survey geometry-based mask (mask d). Furthermore, while the improvements of the value-added catalogue only appear above 4 mJy, it is not suitable for a cosmological analysis requiring a homogeneous sample starting from 2 mJy. Given these findings, we adopt `mask d' and a flux density threshold of at least $2$~mJy as the default selection for the cosmological analysis of LoTSS-DR2.

In the right panel of Fig.~\ref{fig:diff_source_counts} we extend our comparison of `mask d' source counts with results from various radio surveys at different frequencies. Specifically, we compare with the RACS-low source catalogue \citep{hale2021} at 888~MHz, 
the LoLSS-DR1 source catalogue \citep{deGasperin2021} at 54~MHz, 
and the NVSS source catalogue \citep{matthews2021} at 1.4~GHz. 
These three results are scaled to 144~MHz with a spectral index between $-0.8$ and $-0.7$, resulting in bands of differential source counts. 
In addition, we compare with the SKADS catalogue \citep{Wilman2008} at 1.4~GHz, scaled to 144~MHz using $\alpha=-0.7$,
and with the TGSS best-fit polynomial \citep{intema2017}, scaled from 150~MHz using $\alpha=-0.73$ (the median spectral index of \citealp{intema2017}).
It is shown that up to a different flux density scaling of around 10\%, our result agrees very well with previous results. 

A key contribution to the difference between the LoTSS-DR2 and the LoTSS Deep Field differential source counts is that in the deep fields the individual radio components have been identified and combined into single sources. Additionally, flux scale differences exist between the deep fields and LoTSS-DR2.

\section{Results}\label{sec:results}

In the following, we present results for three different flux density thresholds, spaced equally on a logarithmic scale, namely 2, 4, and 8 mJy, resulting in 
$827\,362$, $460\,572$, and $279\,304$ radio sources, respectively. 
Without applying any other thresholds, \citet{Siewert2019} demonstrated that LoTSS-DR1 data above 2 mJy produced robust and reliable estimates of the two-point correlation. An additional cut on the signal to noise ratio (defined as peak flux density over rms noise) allowed  \citet{hale2021} to lower that threshold to 1.5 mJy. Looking at a variety of flux density thresholds, allows us to test the robustness of our findings.
\subsection{Counts-in-cells distribution}\label{sec:countsincellsdistribution}

In this section, building on the discussions in Sect.~\ref{sec: countsincells_theory}, we compare the counts-in-cells model distributions  with the LoTSS-DR2 data. The parameters of these model distributions are determined by means of the method of moments, based on the empirical estimates of the mean $\hat{\mu}$ and variance $\widehat{\sigma^2}$. Table~\ref{tab:Distribution_parameters} presents the parameter values for the Poisson, compound Poisson, and negative binomial distributions, evaluated for mask d and at different flux density thresholds.

\begin{figure*}
  
  \includegraphics[width=0.50\linewidth]{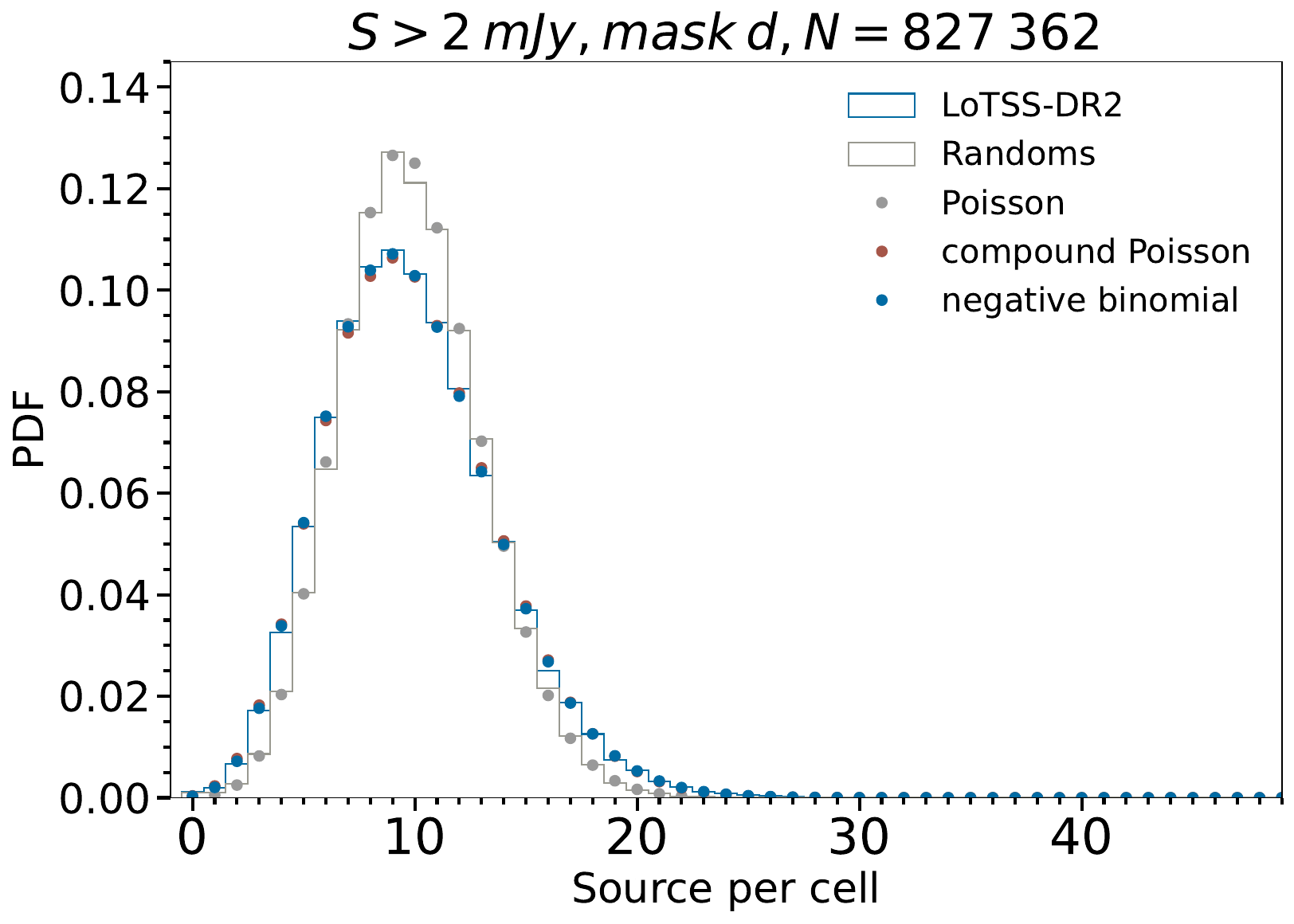}
  \includegraphics[width=0.50\linewidth]{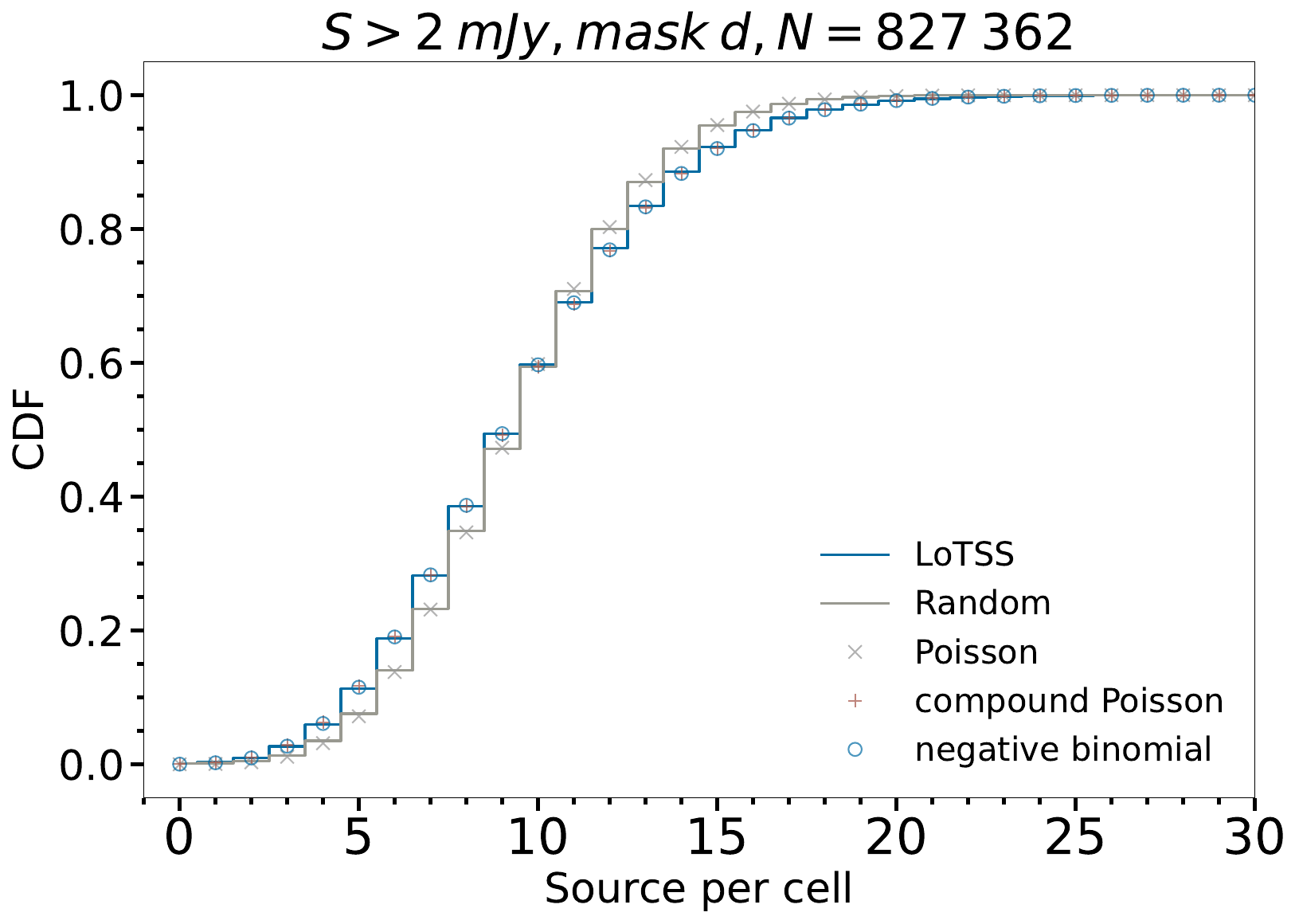}
   
  \includegraphics[width=0.50\linewidth]{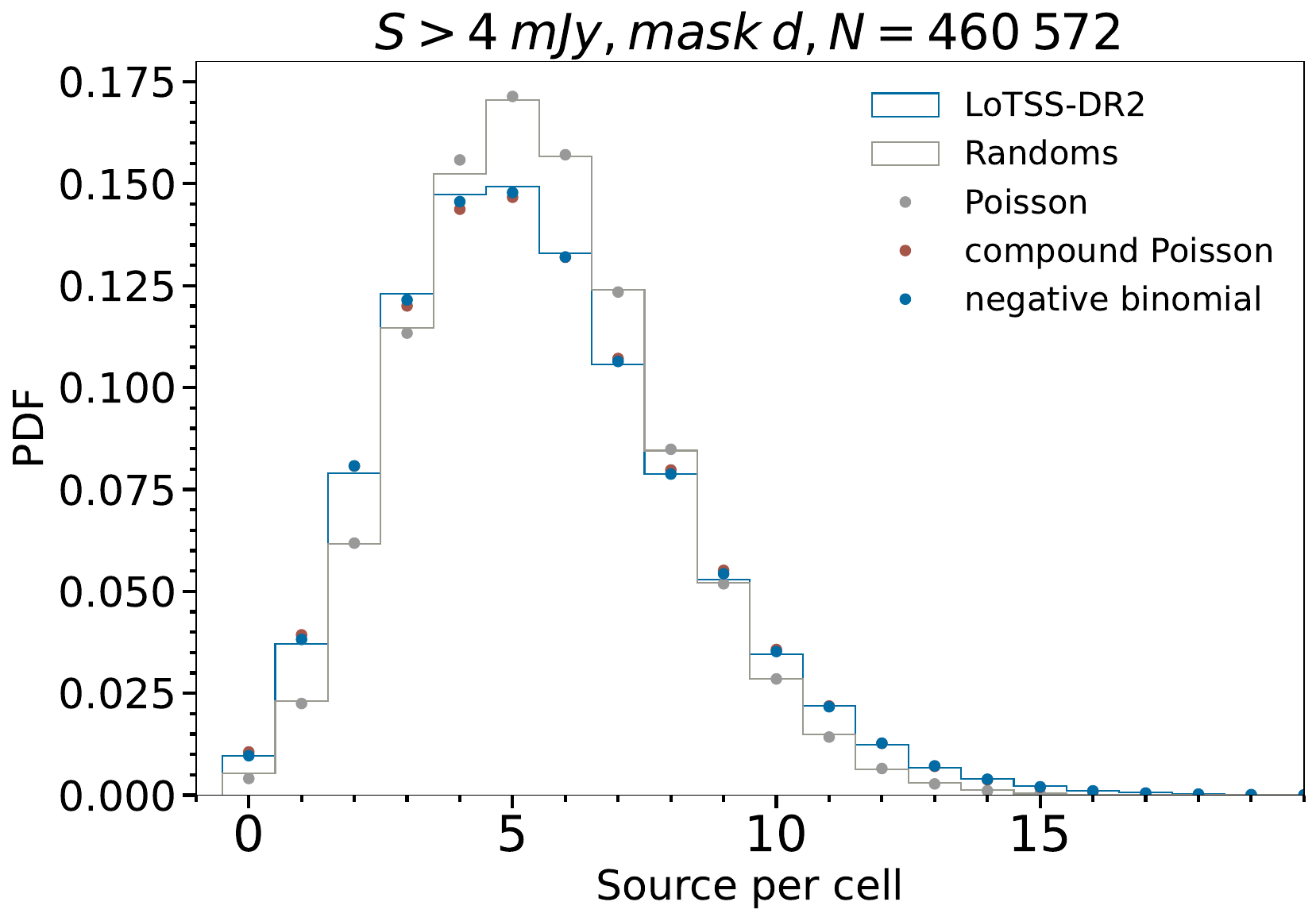}
  \includegraphics[width=0.50\linewidth]{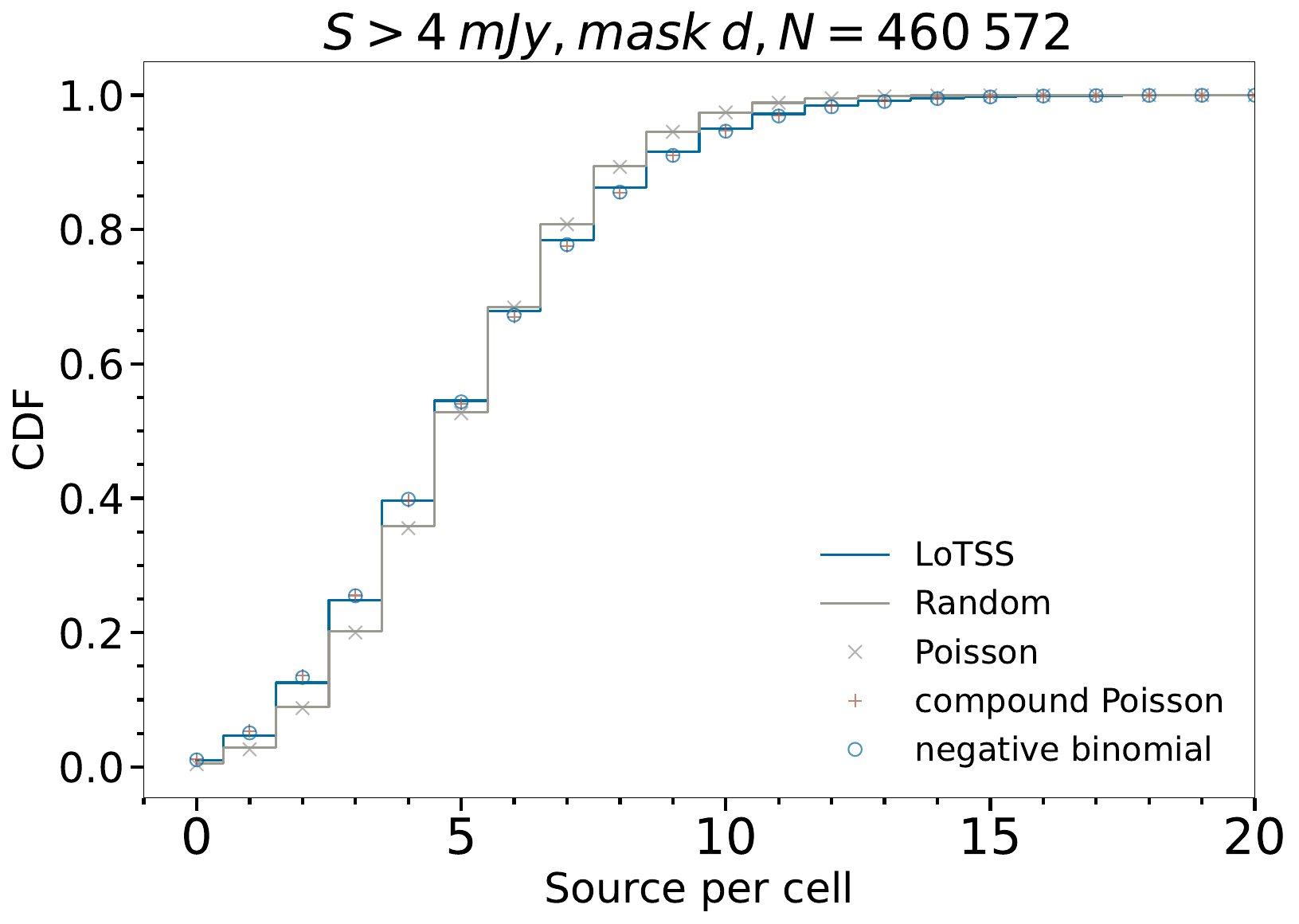}
  
  \includegraphics[width=0.50\linewidth]{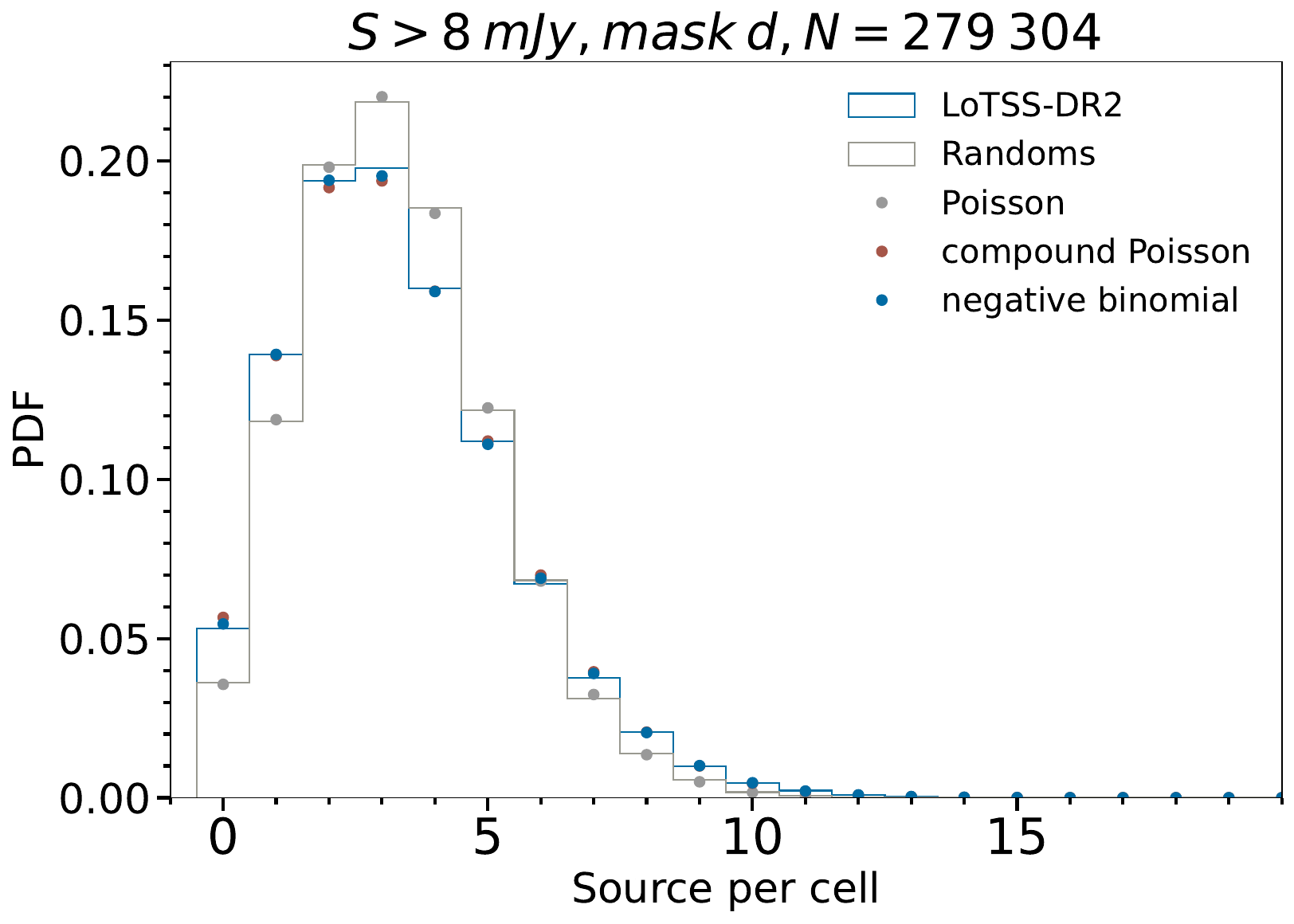}
  \includegraphics[width=0.50\linewidth]{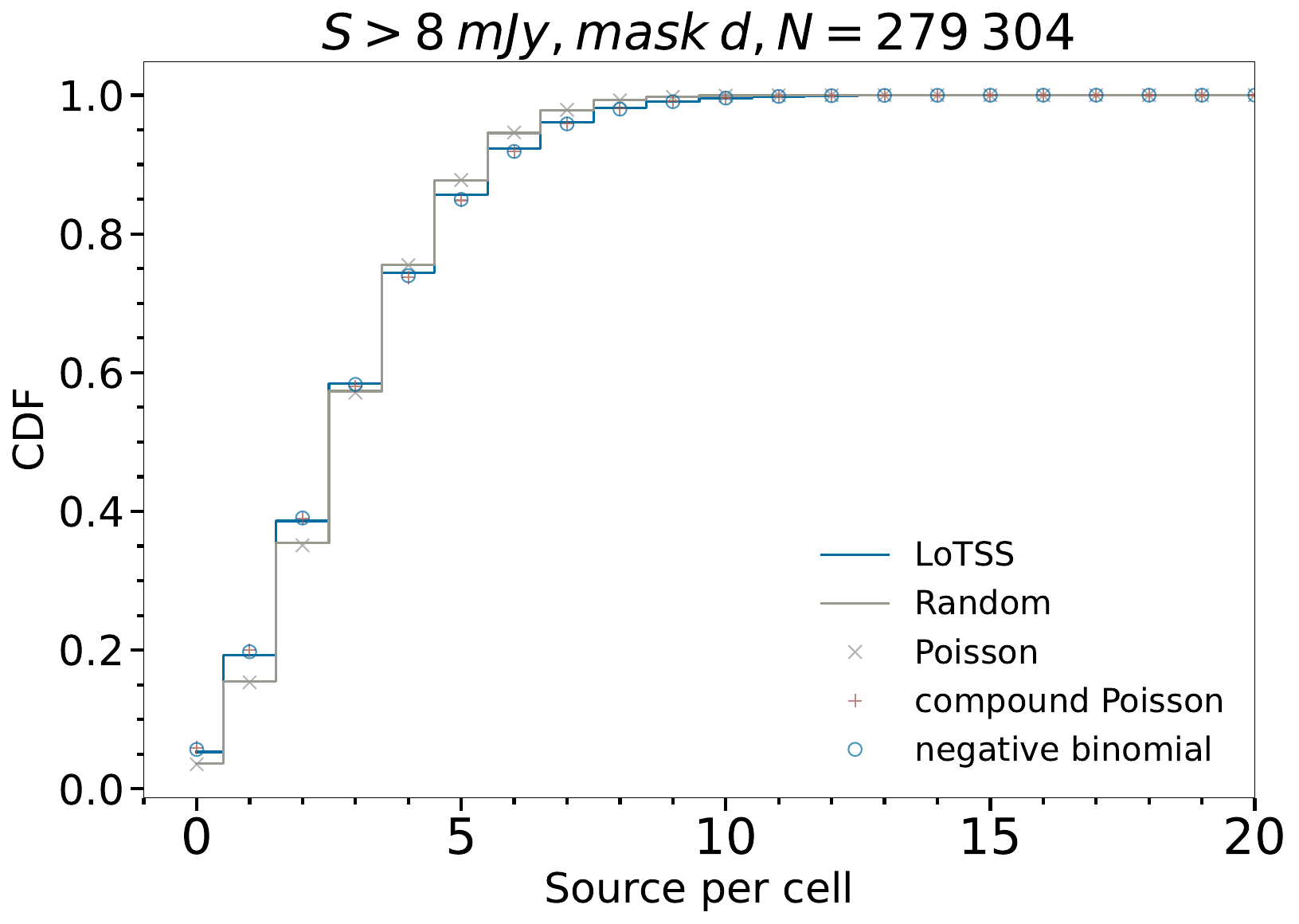}
  
  \caption{Histograms of counts-in-cells of mask d for LoTSS-DR2 and the random mock catalogue at the flux density thresholds 2, 4 and 8~mJy (\textit{left, top to bottom}), and the CDFs (\textit{right, top to bottom}) with the best-fit Poisson and compound Poisson and negative binomial distributions.}
  \label{fig:HistogramDistributions}
\end{figure*}

The left panels of Fig.~\ref{fig:HistogramDistributions} display the corresponding histograms of the counts-in-cells distribution at different flux density thresholds, comparing the LoTSS-DR2 data and the random mock catalogue. Since the mock catalogue contains more sources than the actual data, we randomly selected an equivalent number of sources after applying the same mask and flux density threshold. The observed distribution of sources (blue histograms) deviates from a Poisson distribution (the grey dots). For increasing flux density thresholds, this deviation decreases gradually, but remains clearly visible. In contrast, the random mock catalogue follows a Poisson distribution, with minor deviations arising from survey systematics (see also Sec.~\ref{sec:mock}).

\begin{table}
	\caption{Values of the parameters of three different models for the counts-in-cells distribution of LoTSS-DR2 sources.}
	
    \setlength{\tabcolsep}{4.5pt}
    \renewcommand{\arraystretch}{1} 
    
	\begin{tabular}{ccc|cc|cc}
		\hline \hline
        Mask &
		\multicolumn{1}{p{1.5cm}}{\centering $S_\mathrm{min}$} & 
        \multicolumn{1}{p{1cm}}{\centering Poisson} &
        \multicolumn{2}{p{1cm}}{\centering compound \\ Poisson} &
        \multicolumn{2}{p{1cm}}{\centering negative \\ binomial} \\
		\cmidrule(lr){3-7}
        \addlinespace[0.5ex]
         & &
		\multicolumn{1}{p{1.3cm}}{ \centering $\lambda$} & 
        \multicolumn{1}{p{0.4cm}}{\centering $\lambda$} &
        \multicolumn{1}{p{0.4cm}}{ \centering $\kappa$} &
        \multicolumn{1}{p{0.4cm}}{\centering r} &
        \multicolumn{1}{p{0.4cm}}{\centering p}\\
		\hline
		\addlinespace[0.5ex]
		\multirow{4}{3em}{mask d} 
 
        & 2 mJy & $9.91$ & $17.21$ & $0.58$ & $17.21$ & $0.37$  \\ & 4 mJy & $5.51$ & $12.18$ & $0.45$ & $12.18$ & $0.31$ \\
		& 8 mJy & $3.34$ & $9.63$ & $0.34$ & $9.63$ & $0.26$ \\
		\hline
		
	\end{tabular}
\label{tab:Distribution_parameters}
\end{table}

Based on the measured value of the clustering parameter $n_\mathrm{c}$ (Eq.~\ref{eq:4}), the excess variance in the random mock catalogue is about $10 \%$ at 2 mJy, likely due to various systematic effects incorporated into the randoms. In contrast, the excess variance for the LoTSS-DR2 sources is about $46\%$ which cannot be understood by the effects incorporated in the random mock catalogue (see Fig. \ref{fig:HistogramDistributions}). This supports the hypothesis that the multi-component nature of radio sources plays a key role in shaping the distribution of the counts. However, some of this excess variance may still be due to imaging artefacts. Notably, both the negative binomial (blue dots) and compound Poisson (red dots) fit quite well, though a clear preference is not immediately evident. We make use of both frequentist and Bayesian statistical tests to examine which model describes the distribution of sources better. 

The Pearson $\chi^2$ test \citep{Pearson1900} compares the expected and observed distribution of events sorted into classes or bins, 
\begin{equation}
	\chi^2 = \sum_{i=1}^{n} \frac{(O_i - E_i)^2}{E_i},
\end{equation}
where $O_i$ and $E_i$ are the observed and expected abundances of event $i$, and n is the number of event classes or bins. 
This statistic depends on the number of degrees-of-freedom (dof), which are the number of histogram bins minus the number of fitted parameters of the distribution. This allows us to account for the fact that models with more fitted parameters might fit the data better, simply because they have more flexibility. The so-called reduced chi-square statistic \citep{Wong1992computational}, which is defined as $\chi^2$/dof, is hence used to reduce these dependencies. 
A value of $\chi^2 $/dof of around unity indicates a good fit, meaning that it is likely that the data have been drawn from the proposed model. If the value is much greater than 1, it suggests that the data are unlikely given the model, while a value significantly less than 1 could imply that the model is overfitting the data.

To minimise the impact of outliers, we limit the test to event classes with significant expected and observed abundance (in order to avoid division by small numbers). We limit the analysis to histogram bins containing at least 1000 cells to ensure the histograms are robust, free from distortions and not unduly affected by low-count bins. In Fig.~\ref{fig:HistogramDistributions}, this corresponds to the bins of 3 to 18 sources, 1 to 12 sources and 0 to 8 sources for 2, 4 and 8 mJy flux density threshold, respectively. Bins with fewer than 1\,000 cells, at flux density above 2 mJy, constitute less than 10\% of all sources. Table~\ref{tab:ChiSquared} shows the reduced chi-square test for the mask d and for the different flux cuts. The negative binomial distribution is preferred for all flux density thresholds. This holds also for the other geometry-based masks detailed in App.~\ref{sec:Geometry-based_masks_results}.

\begin{table}
	\caption{$\chi^2$/dof for three models of the counts-in-cells for mask d  at different flux density thresholds for LoTSS-DR2 sources. The degrees-of-freedom are $N_\mathrm{bins} - 1$ for the Poisson distribution and $N_\mathrm{bins} - 2$ for the compound Poisson and negative binomial distributions.}
	\begin{tabular}{c|c|c|c|c}
		\hline \hline
		\multicolumn{1}{p{1.4cm}}{\centering $S_\mathrm{min}$} & 
        \multicolumn{1}{p{1.4cm}}{\centering $N_\mathrm{bins}$} &
        \multicolumn{1}{p{1.4cm}}{\centering Poisson} &
        \multicolumn{1}{p{1.4cm}}{\centering compound \\ Poisson} &
        \multicolumn{1}{p{1.4cm}}{\centering negative \\ binomial} \\
		\hline
		\addlinespace[0.5ex]
		\multirow{3}{*}
        
         2 mJy & $16$ & $319.8$ & $11.7$ & $7.8$ \\ 4 mJy & $12$ & $325.5$ & $10.9$ & $5.2$ \\ 8 mJy & $9$ & $287.1$ & $14.9$ & $6.4$ \\
		\hline
	\end{tabular}
	\label{tab:ChiSquared}
\end{table}

\citet{Pettitt1977} demonstrated that for discrete distributions, the Kolmogorov-Smirnov (KS) test might exhibit higher statistical power compared to the $\chi^2$ test. Therefore, we make use of the KS test as our secondary evaluation method. The KS test is non-parametric, meaning it makes no assumption about the distributions being compared. It quantifies the distance between the empirical cumulative distribution function (CDF) of the sample $F_n(x)$ 
(where the index $n$ stands for the number of data bins) and the CDF of the reference distribution $F(x)$ (see \citealp{lista2017}). The KS statistic, or $d$-value, measures the maximum vertical distance between the two CDFs:
\begin{equation}
	d_n = \underset{x}{\sup} \:|F_n(x) - F(x)|.
\end{equation}
The null hypothesis, namely that $F_n$ is a random realisation of the model distribution $F$, is rejected at confidence level $(1-\alpha)$ if $d_n > d_\alpha$, where $\alpha$ is the frequentist probability for a false rejection of the null hypothesis. For details on the critical values $d_\alpha$, refer to \cite{FellerKSTest1948} and \cite{Smirnov1948}.

However, when the form or the parameters of the model distribution $F(x)$ are estimated from the data, the tabulated critical values provided in standard references no longer apply. In this case, Monte Carlo methods must be used to determine the critical values. We thus simulate realisations of the hypothesised distribution and calculate the 99\% confidence level (C.L.) of the resulting value of $d_\alpha$, which serve as our critical values for rejecting the null hypothesis that the theoretical distribution matches the observed data.

\begin{table*}
\centering	\caption{Kolmogorov-Smirnov test statistic for mask d at different flux density thresholds of the LoTSS-DR2 sources. \(d_n\) denotes the measured test statistic and $ d_{\alpha}$ its critical value at 99\%  confidence level. 
 }
	\begin{tabular}{cccc|cc|cc}
      \hline \hline
        &
		\multicolumn{1}{p{1.5cm}}{\centering $S_\mathrm{min}$} & 
        \multicolumn{2}{p{2.5cm}}{\centering Poisson} &
        \multicolumn{2}{p{2.cm}}{\centering compound \\ Poisson} &
        \multicolumn{2}{p{2cm}}{\centering negative \\ binomial} \\
		\cmidrule(lr){3-8}
        \addlinespace[0.5ex]
         & &
		\multicolumn{1}{p{1.3cm}}{ \centering $d_n$} & 
        \multicolumn{1}{p{1.cm}}{\centering $d_{\alpha}$} &
        \multicolumn{1}{p{1cm}}{ \centering $d_n$} &
        \multicolumn{1}{p{1cm}}{\centering $d_{\alpha}$} &
        \multicolumn{1}{p{1cm}}{\centering $d_n$} & 
        \multicolumn{1}{p{1.cm}}{ \centering $d_{\alpha}$}\\

		\hline
		\addlinespace[0.5ex]
        & 2 mJy & $0.0561$ & $0.0033$ & $0.0102$ & $0.0031$ & $0.0078$ & $0.0030$ \\ 
        & 4 mJy &$0.0507$ & $0.0030$ & $0.0080$ & $0.0028$ & $0.0052$ & $0.0029$ \\
		& 8 mJy & $0.0407$ & $0.0030$ & $0.0064$ &  $0.0028$ & $0.0049$ & $0.0027$\\
		\hline
		\addlinespace[0.5ex]
	\end{tabular}
	\label{tab:KStest_table}
\end{table*}

The CDFs of the counts-in-cells for the data, random mocks and three theoretical models are shown on the right-hand panels of Fig.~\ref{fig:HistogramDistributions}. 
Table~\ref{tab:KStest_table} presents the measured KS test statistic \(d\)-value and its critical values $d_{\alpha}$ at a significance level of $\alpha = 0.01$ for various flux density thresholds. All \(d\)-values for the Poisson distribution at all flux density thresholds exceed the critical values by a large margin, leading to the rejection of the Poisson distribution with very high confidence ($> 99\%$ C.L.). This is consistent with the results drawn from the reduced chi-square test discussed above. Similarly, for the compound Poisson distribution, the measured \(d\)-values are two or three times larger than the critical values, allowing us to exclude this distribution across all flux density thresholds. For both the Poisson and compound Poisson distributions, we used 1000 Monte Carlo realisations to calculate the critical \(d\)-values. Given that the measured values were already significantly larger, extreme precision in handling outliers was unnecessary. However, for the negative binomial case, where the measured and critical $d$-values are closer to each other, we increased the number of realisations to 10\,000 to account for potential outliers and ensure greater accuracy. For the negative binomial distribution with mask d, the lower observed \(d\)-values compared to those of the compound Poisson distribution suggest that the negative binomial distribution is preferred. However, the KS test still rejects this distribution, indicating that other effects on top of the ones accounted for in the negative binomial distribution must exist. 

Both Pearson's chi-square statistic and the KS test clearly exclude a Poissonian distribution of the data, while the random mocks are in good agreement with a Poissonian distribution. Furthermore, both tests favour the negative binomial distribution 
over a compound Poissonian distribution. To further quantify that preference, we also embark on a Bayesian model comparison by computing the Bayes factor ($B$). The Bayes factor, which is the ratio of the evidence of two models, quantifies the support for one model over the other. Essentially, it is the ratio of the posterior odds to the prior odds, and if the priors for both models are the same, it reduces to the posterior odds.

We calculate the Bayes factor for the hypothesis of the negative binomial over the compound Poisson distribution. When the prior odds are presumed to be non-informative, the posterior odds reduce to the ratio of their likelihoods, which defines the Bayes factor (\citealp{jeffreys1961,gelman2003bayesian}) 
\begin{equation} \label{bayes_factor}
    B_{\mathrm{NB-CP}} = \frac{p(\mathrm{data}\,|\,\mathrm{model}_{\mathrm{NB}})}{p(\mathrm{data}\,|\,\mathrm{model}_{\mathrm{CP}})}.
\end{equation}

For flux density thresholds of 
2, 4, and 8 mJy and mask d, the Bayes factors are 25.4, 17.2, and 15.9, respectively. According to Jeffrey's scale of evidence (\citealt{jeffreys1961}), a Bayes factor higher than 10 provides strong evidence for a model. Therefore, 
there is strong evidence in favour of the negative binomial model.

We conclude that the negative binomial distribution provides an excellent fit to the data and offers a proper physical interpretation especially when using conservative masks (see also App. \ref{sec:Geometry-based_masks_results}). For mask d, our default in the cosmological analysis of LoTSS-DR2, there are some additional effects that are obviously not described perfectly by the negative binomial distribution, since $d_n$ is still larger than $d_\alpha$, while the more aggressive masks discussed in the appendix result in $d_n \sim d_\alpha$ for the negative binomial distribution. Notably, the negative binomial distribution ensures that the number of components starts from at least one, aligning with the expectation of a complete catalogue. This consistency supports the survey's completeness estimates above the 2 mJy flux density threshold.

\begin{figure}
    \includegraphics[width=0.98\linewidth]{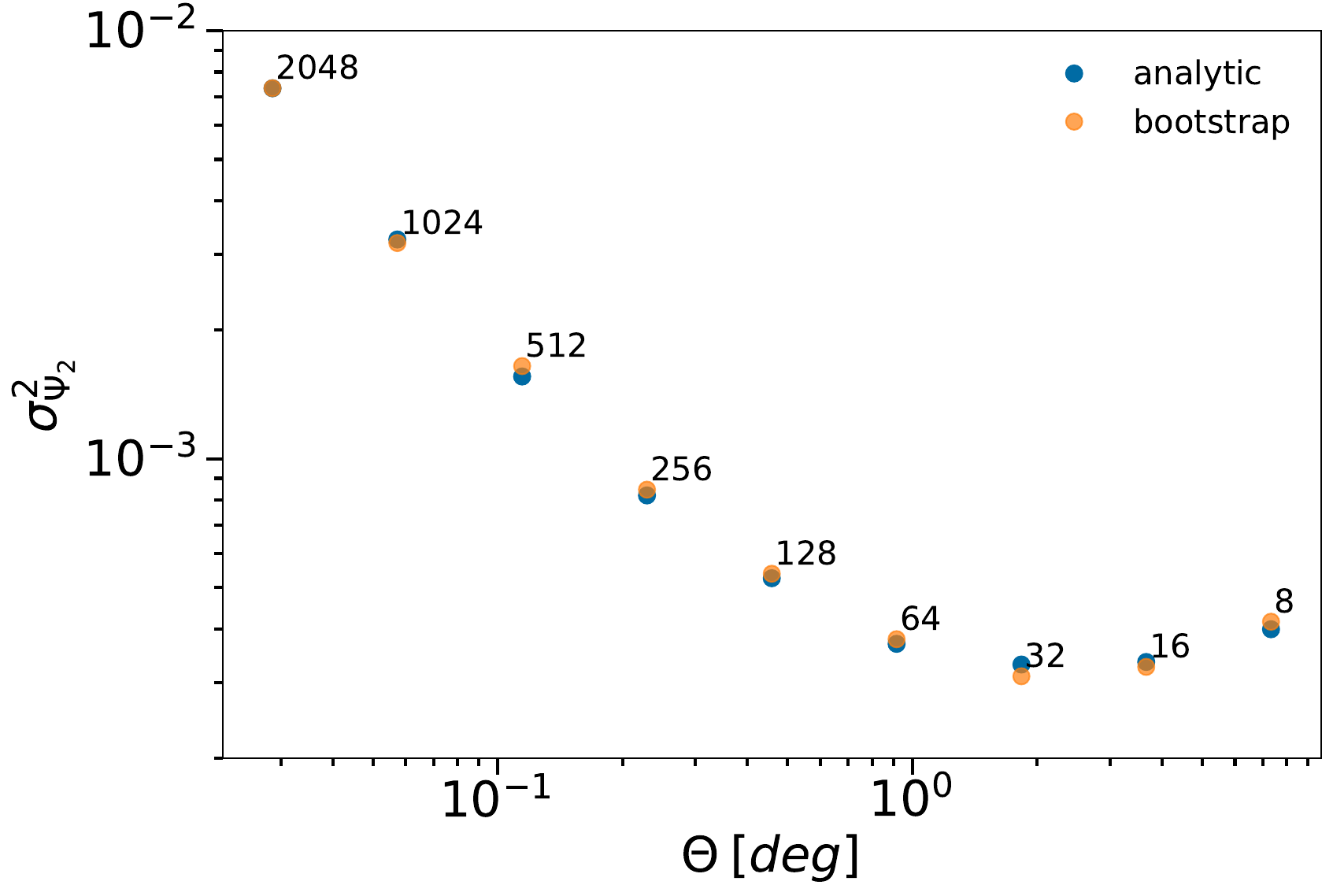}
    \caption{Empirical variance of $\Psi_2$ estimated analytically (blue) and from a bootstrap method (orange) with 500 realisations, each at 2 $\mathrm{mJy}$ flux density threshold at different values of $N_\mathrm{side}$ of the {\sc HEALPix} scheme.}
    \label{fig:Error_psi}
\end{figure}

\begin{figure*}
    \centering
    \includegraphics[width=0.49\linewidth]{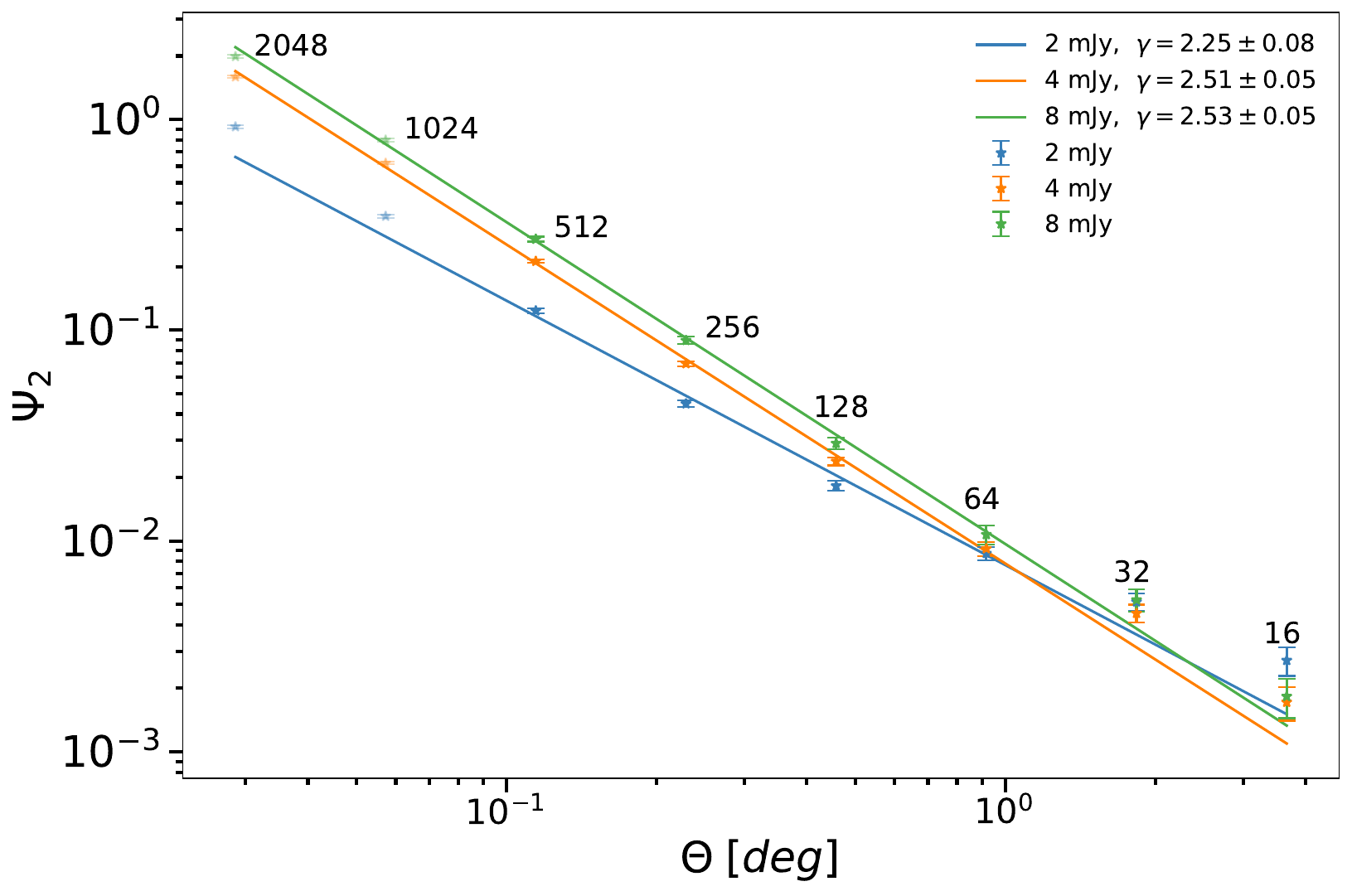}
    \includegraphics[width=0.49\linewidth]{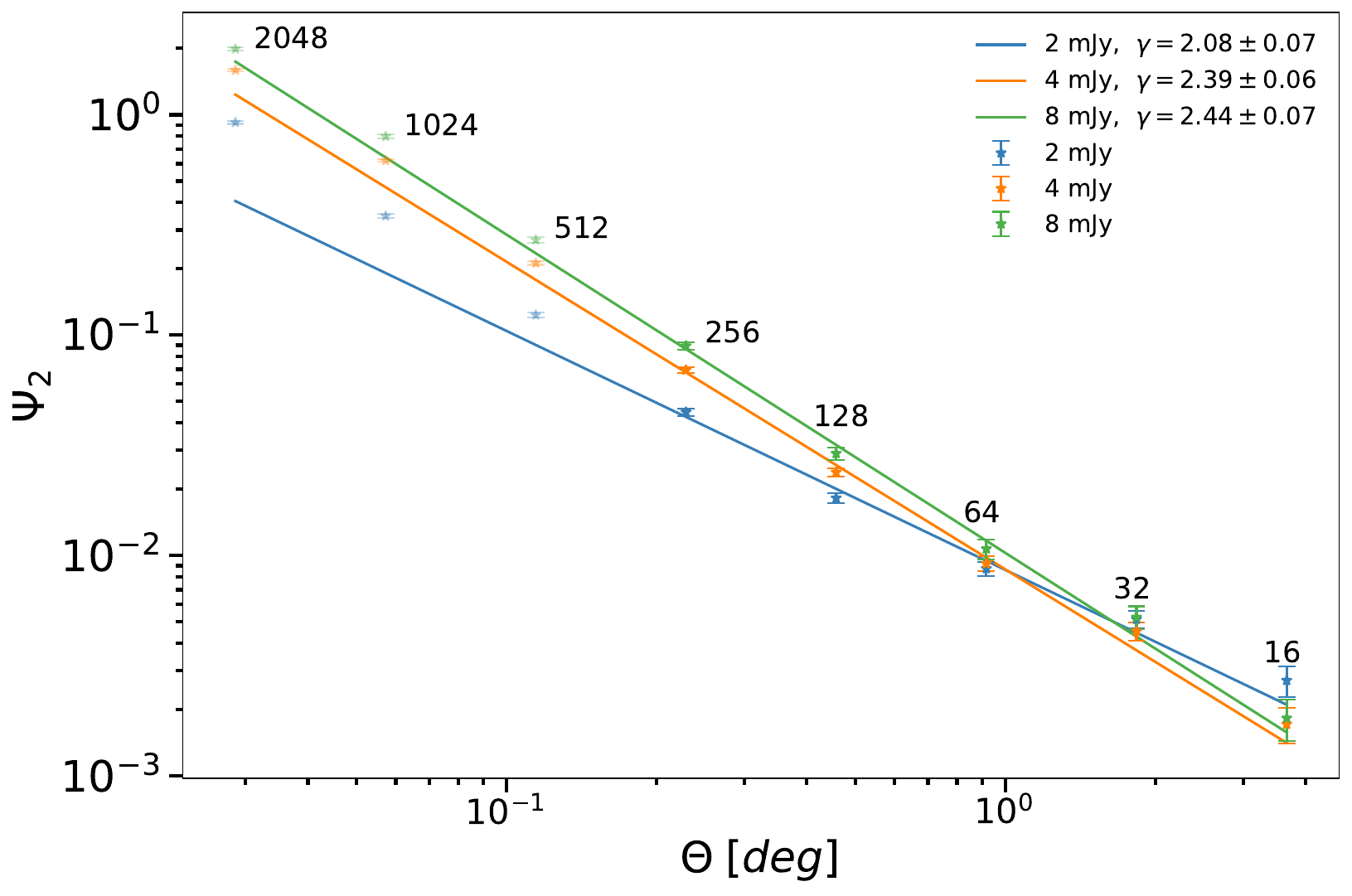}
    \caption{ Fitted single power law for variance variations of re-scaling HEALPix maps for $N_\mathrm{side}=16-512$ corresponding to $3.8$~deg to $0.11$~deg at different flux density thresholds (\textit{left}) and for $N_\mathrm{side}=16-256$ corresponding to $3.8$~deg to $0.23$~deg (\textit{right}), plotted for $N_\mathrm{side}=16-2048$, which means the fit was not performed for the fainter data points.}
    \label{fig:fitnside16_152_256}
\end{figure*}

\begin{figure*}
    \centering
    \includegraphics[width=0.49\linewidth]{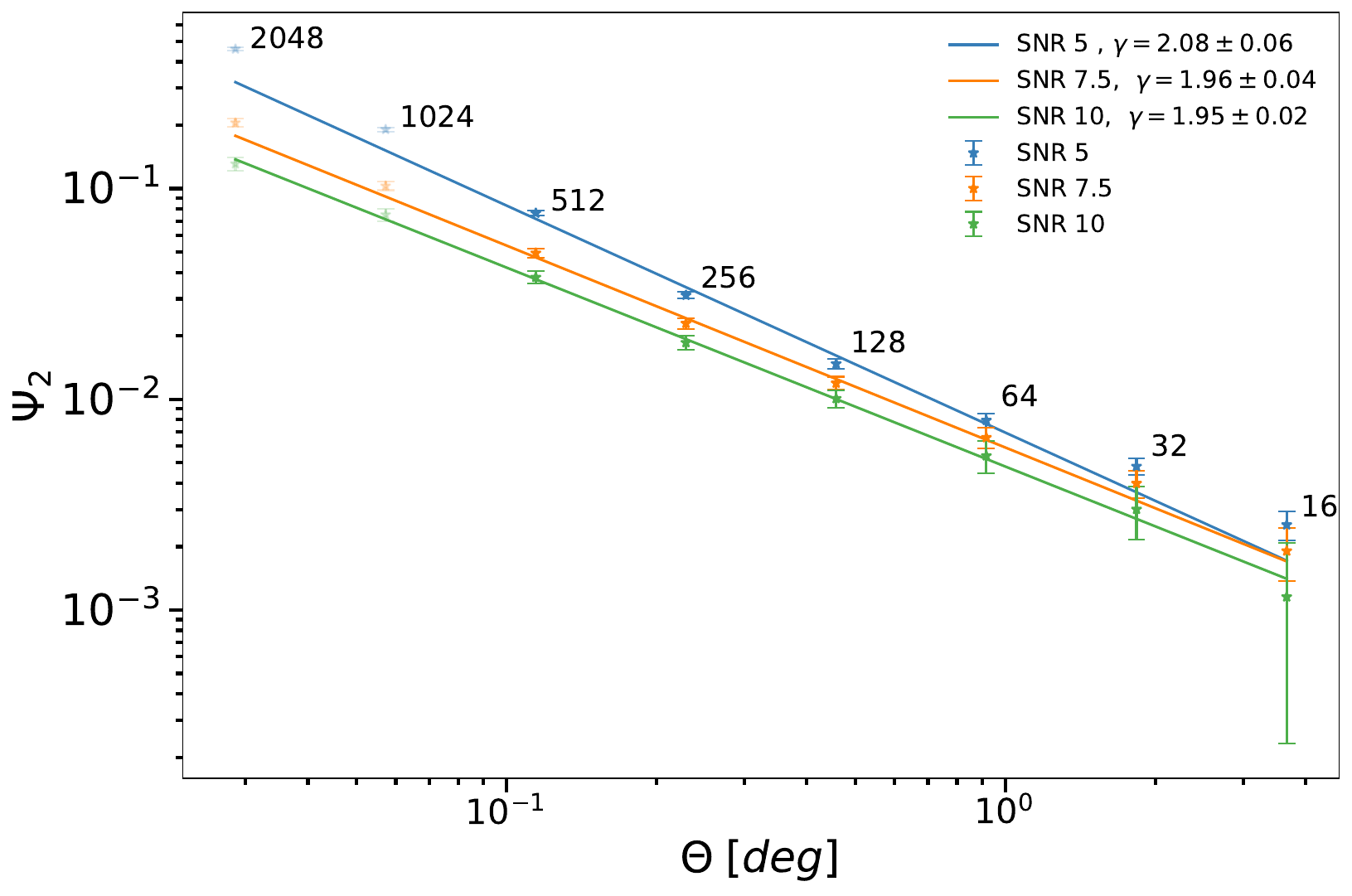}
    \includegraphics[width=0.49\linewidth]{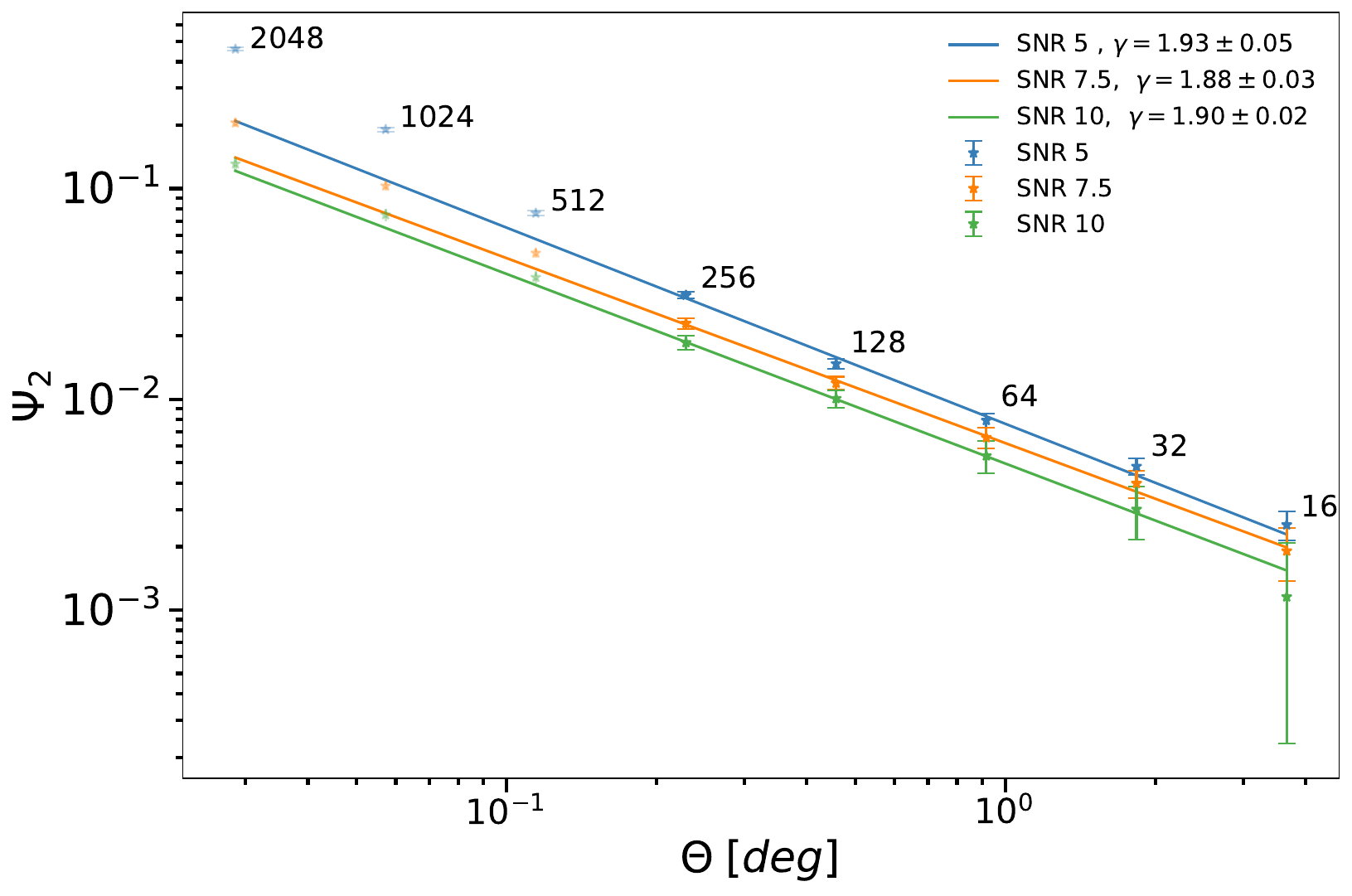}
    \caption{Fitted single power law for variance variations of re-scaling HEALPix maps for $N_\mathrm{side}=16-512$ (\textit{left}) and for $N_\mathrm{side}=16-256$ (\textit{right}) at flux density threshold 2 mJy and different SNR cuts plotted for $N_\mathrm{side}=16 - 2048$ (solid points represent the data points for which fits were performed).}
    \label{fig:scaling_fit_F2diffSNR}
\end{figure*}

\begin{figure}
    \includegraphics[width=\linewidth]{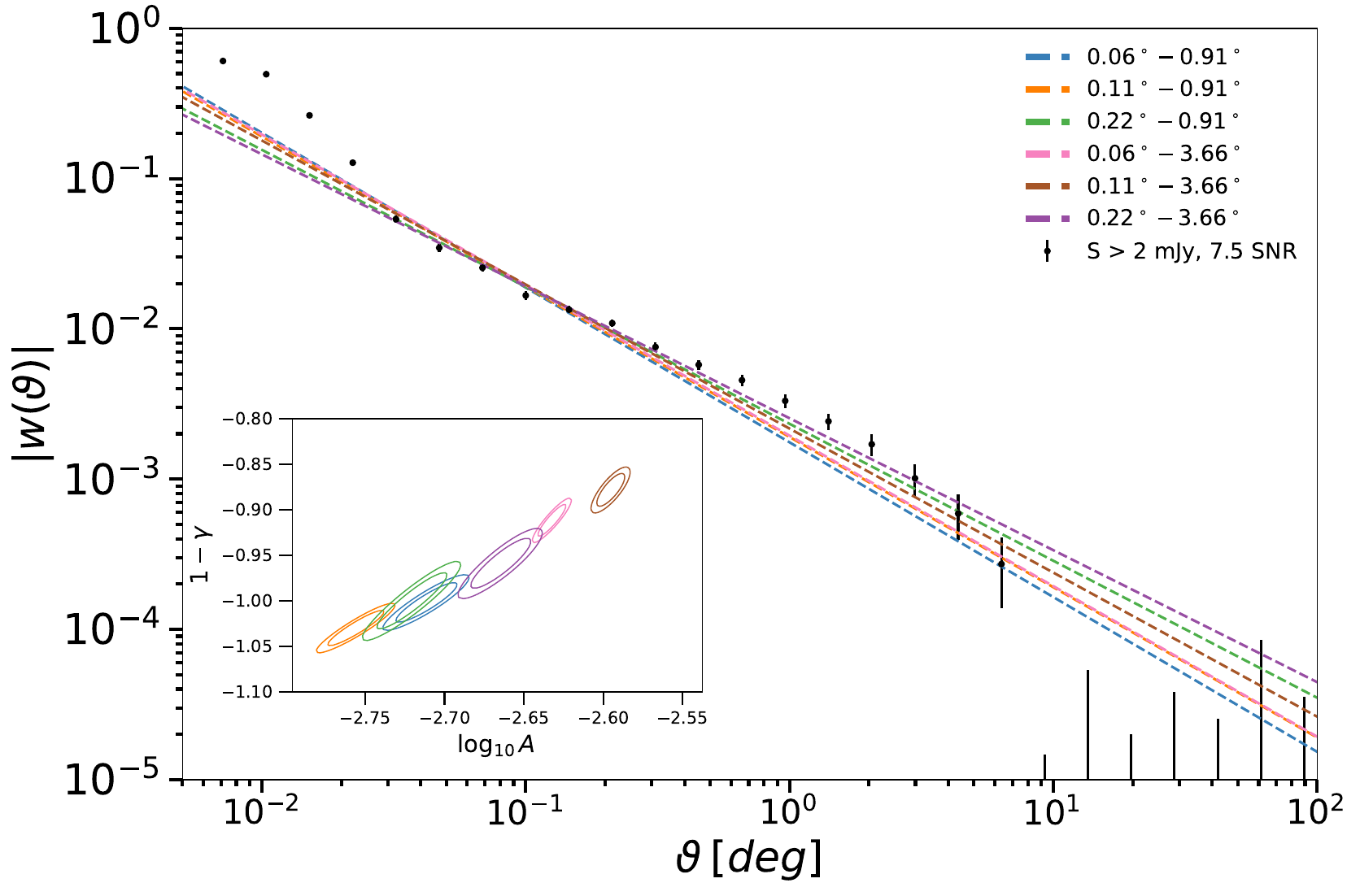}
    \caption{Fitting results for power laws to different angular ranges at 2 mJy flux density threshold and SNR 7.5. (dashed lines). The dots represent the results of direct measurements of the two-point correlation function using the Landy-Szalay estimator. The inner plot shows the $1 \sigma$ and $2 \sigma$ contours for amplitude and exponent of the power-law fit to $w(\vartheta)$.}
    \label{fig:TPCF}
\end{figure}

\subsection{Angular correlation function from scaling of cell size} \label{sec: AngularCorrelationFunction}

As detailed in Sect.~\ref{sec:scalingofcellsizestheory}, the reduced normalised variance $\Psi_2$ (Eq.~\ref{eq:23}) of counts-in-cells is directly linked to the angular two-point correlation function. Through fitting this statistic over various cell sizes to a power law ansatz for the angular two-point correlation function, we can determine both the amplitude and exponent of the power law. To account for uncertainties, we employ uncertainty propagation, and in addition, we use the bootstrap method. 
To estimate the uncertainty of $\Psi_2$, we propagate the uncertainties,
 \begin{equation}\label{eq27}
     {\sigma}^2_{\Psi_2} = \left(\frac{\partial \Psi_2}{\partial m_2}\right)^2 \! {\sigma}^2_{m_2} + \left(\frac{\partial \Psi_2}{\partial \mu}\right)^2 \! {\sigma}^2_\mu + 2 \frac{\partial \Psi_2}{\partial m_2} \frac{\partial \Psi_2}{\partial \mu} \sigma^2_{m_2, \mu},
 \end{equation}
where ${\sigma}^2_{\Psi_2}$, ${\sigma}^2_{m_2}$ and ${\sigma}^2_\mu$
are the variances of $\Psi_2$, $m_2$ and $\mu$, respectively and $\sigma^2_{m_2, \mu}$ denotes the covariance of $m_2$ and $\mu$. 
Evaluating the partial derivatives and inserting  
\begin{equation}    
    \sigma^2_\mu = \frac{m_2}{n},
    \sigma^2_{m_2} = \frac{m_4 - (m_2)^2}{n} + \frac{2m_2^2}{n(n-1)},
    \sigma^2_{m_2, \mu} = \frac{m_3}{n},
\end{equation}
from \citet{Shanmugam2008} in Eq.~(\ref{eq27}), we get
\begin{equation}
    {\sigma}^2_{\Psi_2}\!\! = \!\frac{1}{n\mu^4}\! \left[
    m_4\! + \!2m_3\! + \!m_2\! +\! \frac{4m_3^2} {\mu^2}\! - \!\frac{n\!-\!3}{n\!-\!1} m_2^2
   \! -\! \frac{4m_2(m_3\! +\!m_2)}{\mu}\right].
\end{equation}
This quantity can then be estimated by means of the observed moments $\hat{\mu}, \hat{m}_2, \hat{m}_3$ and $\hat{m}_4$.

We also employ the bootstrap method (\citealt{Bootstrap}) and compare its results with those obtained from uncertainty propagation. The bootstrap is a resampling technique that generates multiple datasets by randomly selecting samples from the original dataset,  with the same size as the original dataset, allowing for repeated selection of the same data points. This method is widely used to estimate and visualize the sampling distribution of a statistic. The bootstrap-generated distribution can be used to compute standard errors and confidence intervals for statistical estimates. One key advantage of the bootstrap method is that it does not require model assumptions about the underlying data distribution or the normality of the sample statistic, making it particularly useful in cases where traditional parametric methods may not be applicable.

In Fig.~\ref{fig:Error_psi}, we present the results of error analysis using both analytical uncertainty propagation and the bootstrap method at a flux density threshold of 2 mJy. Notably, the comparison reveals a high degree of consistency between the two approaches.

In the following, we determine $\Psi_2(\Theta)$, and fit a power-law parametrisation of the angular two-point correlation. After numerically computing the coefficients $C_{\gamma}$ (as detailed in App.~\ref{sec:cgamma}) and subtracting the $\Psi_2(\Theta)$ values obtained from the random catalogue, we fit a power law to derive parameters $A_0$, and $\gamma$.  To achieve this, we manipulate mask d by both upscaling and downscaling, then apply it to individual {\sc HEALPix} maps at varying $N_\mathrm{side}$ resolutions. Our calculations encompass the computation of the reduced normalised variance across $N_\mathrm{side}$ values ranging from 16 to 512, covering angular scales spanning from $3.66$~deg down to $0.11$~deg. Additionally, we perform this analysis over $N_\mathrm{side}$ values from 16 to 256, corresponding to angular scales ranging from $3.66$~deg down to $0.22$~deg. Below $0.1$~deg, multi-component source clustering becomes important (see also \citealt{Hale2023}).

The fitting parameters and results of the reduced chi-square test at different flux density thresholds are presented in Table~\ref{tab:fitting_results_diffFlux}. Fig.~\ref{fig:fitnside16_152_256} illustrates the best fit values for $\gamma$ across varying flux density thresholds for the two fitted angular distance ranges ($3.66 - 0.11$~deg and $3.66 - 0.22$~deg), with the results extrapolated down to $0.03$~deg.

As observed, higher flux density thresholds result in a decrease in amplitude and an increase in the slope of the power law, indicating an improved fit at these levels. This trend aligns with the findings from LoTSS-DR1 (\citealp{Siewert2019}; \citealp{Nitesh2024}) but contrasts with results from \citet{Wilman2003} for the Boötes Deep Field and \citet{Rana&Bagla2019} for TGSS-ADR1. This discrepancy may stem from flux scale calibration issues in the LoTSS pipeline, as previously discussed by \citet{Shimwell2022} and \citet{Hale2023}. Alternatively, it could reflect a shift in the distribution of source populations with flux density, where a larger fraction of star-forming galaxies (SFGs) dominates at lower flux density thresholds (\citealp{DeepFields_Best2023}). At higher thresholds, the clustering signal weakens due to fewer components being associated with each source. The steeper slope at these flux densities is likely driven by the increasing dominance of active galactic nuclei (AGNs), which tend to exhibit lower clustering amplitudes (\citealp{Magliocchetti2016}; \citealp{Hale2018}).

\begin{table}
	\caption{Best-fit results for the amplitude and exponent of the power-law ansatz for the angular two-point correlation at different flux density thresholds, shown for two $N_\mathrm{side}$ ranges: 16 -- 256 and 16 -- 512. $C_{\gamma}$ is computed for each $\gamma$, and the reduced chi-square is provided for each fit, along with the corresponding number of degrees of freedom.} 
       \setlength{\tabcolsep}{3.2pt} 
    \renewcommand{\arraystretch}{1}
	\begin{tabular}{ccccccc}
		\hline \hline
		\addlinespace[0.5ex]
        $N_{side}$ & 
		\multicolumn{1}{p{0.05ex}}{\centering $S_{\mathrm{min}}$} & 
        \multicolumn{1}{p{1.5cm}}{\centering $\gamma$} &
        \multicolumn{1}{p{0.5cm}}{\centering $C_{\gamma}$} &
        \multicolumn{1}{p{2.5cm}}{\centering $A$} &
        \multicolumn{1}{p{0.5cm}}{\centering $\frac{\chi^2}{\mathrm{dof}}$} & 
        \multicolumn{1}{p{0.5cm}}{\centering  dof} \\
		\hline
		\addlinespace[0.5ex]
		\multirow{3}{2.5em}{16 - 256} 
        &2 mJy   &   $2.08 \pm 0.07$  & $3.28$ & $0.0026 \pm 0.0002$ &$3.4$ & 3 \\ & 4 mJy & $2.39 \pm 0.06$  & $5.69$ & $0.0015 \pm 0.0001$ & $2.9$ & 3 \\
		& 8 mJy & $2.44 \pm 0.07$   & $6.29$ & $0.0016 \pm 0.0001$ &$2.1$ & 3\\
		\hline
		\addlinespace[0.5ex]
		\multirow{3}{2.5em}{16 - 512} 
        &2 mJy   &   $2.25 \pm 0.08$  & $4.36$ & $0.0018 \pm 0.0002$ & $8.4$ & 4  \\ & 4 mJy & $2.51 \pm 0.05$  & $7.29$ & $0.0011 \pm 0.0001$ & $5.2$ & 4 \\
		& 8 mJy & $2.52 \pm 0.05$   & $7.51$ & $0.0013 \pm 0.0001$ & $2.5$ & 4 \\
		\hline
		
	\end{tabular}
	\label{tab:fitting_results_diffFlux}
\end{table}

Drawing from the completeness discussions in  \citet{Hale2023}, Section 3.2.3, we examine the fitting results for different signal-to-noise ratios ‘SNR’, defined as peak flux density divided by the rms noise, with the integrated flux density threshold set to 2 mJy (see the fitting results in Table~\ref{tab:fitting_results_diffSNR}). As shown in Fig.~\ref{fig:scaling_fit_F2diffSNR}, there is a slight decreasing trend in the slope as the SNR thresholds increase. Notably, these fits exhibit a closer alignment with lower $N_\mathrm{side}$, which correspond to larger spatial scales. Higher SNRs tend to result in better fits. Table~\ref{tab:fitting_results_diffSNR} contains the fitting parameters and the reduced chi-square test results for different SNRs at a flux density of 2 mJy. As the SNR increases, the amplitude slightly decreases and the slope of the power law decreases, an inverse trend of its power to what we saw for the higher flux density thresholds. Higher SNR results in a decrease in both the slope and amplitude of the correlation function, as brighter, more isolated sources are less clustered than fainter, unresolved ones. Additionally, the potential for overfitting or too large weights, indicated by small reduced chi-square values, suggests that caution should be taken in interpreting the fits, especially when fitting models to high-SNR data.

The results of the parametrised two-point correlation function by fitting power laws to different angular ranges for 2 mJy flux density threshold and SNR 7.5 are presented in Fig.~\ref{fig:TPCF}. We fit power laws to various combinations of $N_{\text{side}}$ values ranging from 16 to 1024, corresponding to angular sizes ranging from $3.66$ - $0.06$~deg.  
Additionally, we estimate the model independent two-point correlation function using the Landy-Szalay estimator \citep{LandySzalay1993} at the same flux density threshold and SNR, utilising the random mock catalogue. The Landy-Szalay estimator is a widely used method for directly measuring the two-point correlation function by comparing the spatial distribution of observed sources to that of a randomly distributed catalogue. This estimator has been shown to have minimal variance and be less biased than other estimators \citep{LandySzalay1993}. To measure it, we use the package {\sc{TreeCorr}} (\citealp{Jarvis2015}) following the parameter settings detailed in Sect.~3 of \cite{Hale2023}. Given the extensive area covered by LoTSS-DR2, we set the separation metric in {\sc{TreeCorr}} to {\textit{`Arc'}} for accurate great circle distance calculations. In addition, we set \textit{bin\_slop} to 0 for precise pair counts within each angular separation bin. 

We also calculated the integral constraint, which accounts for the finite survey area and estimator bias (see e.g. \citealt{Roche1999} and \citealt{siewert2021a}). Since the area of LoTSS-DR2 is much larger than that of LoTSS-DR1, the integral constraint effect is on the order of $10^{-6}$ and is therefore negligible.
 
As clearly shown in Fig.~\ref{fig:TPCF}, the results based on the reduced normalised variance of source counts are in very good agreement with the direct measurements from the Landy-Szalay estimator, particularly at angular scales ranging from $0.3$ down to $0.02$~deg, corresponding to $N_{\text{side}}$ values from 128 to 2048. This agreement validates our approach, demonstrating that the counts-in-cells method provides a reliable alternative for estimating the clustering properties of radio sources. Moreover, our results are consistent with previous studies (e.g., \citealt{Magliocchetti1999}, \citealt{Lindsay2014}), which have reported slopes typically ranging from -1.2 to -0.8 at different flux density limits, further reinforcing the robustness of our findings.

\begin{table}
	\caption{Best-fit results for amplitude and exponent of a power-law ansatz for the angular two-point correlation for 2 mJy flux density threshold at different SNR cuts, for two $N_\mathrm{side}$ ranges 16 -- 256 and 16 -- 512. $C_{\gamma}$ is computed for each $\gamma$, and the reduced chi-square is provided for each fit, along with the corresponding number of degrees of freedom.}
    \setlength{\tabcolsep}{3.5pt} 
    \renewcommand{\arraystretch}{1}
	\begin{tabular}{ccccccc}
		\hline \hline
		\addlinespace[0.5ex]
        $N_{side}$ & 
		\multicolumn{1}{p{.5cm}}{\centering SNR} & 
        \multicolumn{1}{p{1.5cm}}{\centering $\gamma$} &
        \multicolumn{1}{p{0.5cm}}{\centering $C_{\gamma}$} &
        \multicolumn{1}{p{2.5cm}}{\centering $A$} &
        \multicolumn{1}{p{0.5cm}}{\centering $\frac{\chi^2}{dof}$} & 
        \multicolumn{1}{p{0.5cm}}{\centering  $dof$} \\
		\hline
		\addlinespace[0.5ex]
		\multirow{3}{2.5em}{16 - 256 } & 5 &   $1.93 \pm 0.04$  & $2.60$ & $0.0029 \pm 0.0001$ & $1.5$ & 3  \\ &7.5   &   $1.88 \pm 0.02$  & $2.42$ & $0.0025 \pm 0.0007$ &$0.2$ & 3 \\ & 10 & $1.90 \pm 0.02$  & $2.49$ & $0.0018 \pm 0.0005$ & $0.1$ & 3 \\
		\hline
		\addlinespace[0.5ex]
		\multirow{3}{2.5em}{16 - 512} & 5 &   $2.08 \pm 0.06$  & $3.21$ & $0.0022 \pm 0.0002$ & $6.45$ & 4  \\ & 7.5   &   $1.96 \pm 0.04$  & $2.70$ & $0.0021 \pm 0.0001$ & $0.93$ & 4  \\ & 10 & $1.94 \pm 0.02$  & $2.65$ & $0.0018 \pm 0.0001$ & $0.14$ & 4 \\
		\hline
		
	\end{tabular}
	\label{tab:fitting_results_diffSNR}
\end{table}


\section{Discussions} \label{sec:discussion}

\subsection{Negative binomial distribution} \label{NegativeBinomialDistribution}

In this section, we discuss the results of Sect.~\ref{sec:countsincellsdistribution}, where we demonstrated that the negative binomial distribution is the preferred model. At a 2 mJy flux density threshold, the mean number of components per source derived from the negative binomial distribution is \(1.27\) (Eq.~(\ref{eq:nb_components}) and Table~\ref{tab:Distribution_parameters}). This suggests that most sources are likely single entities, possibly unresolved AGNs or star-forming galaxies (SFGs). From the probability distribution function of the logarithmic distribution describing the components of the negative binomial distribution, we find that approximately 80\% of sources have one component, 15\% have two components, and 2\% have three components. These results are consistent with the findings of \cite{bohme2023}, who demonstrated that in the cross-match of the low-resolution (45'') preliminary release of the LOFAR LBA Sky Survey (LoLSS-PR) with LoTSS-DR2 in the HETDEX spring field, the mean number of association components is around 1.33. 
Furthermore, Fig.~8 of \citealt{bohme2023} shows the fraction of sources with different numbers of components, which aligns well with our findings. However, the mean number of components from our results is slightly higher than the mean association component derived from the value-added catalogue (\citealp{hardcastle2023}) at the same flux density threshold, which is $1.13$. The difference likely arises because the value-adding process starts only above 4 mJy. Since the number of sources above 2 mJy is almost double that of sources above 4 mJy (see Fig. \ref{fig:HistogramDistributions}), the mean number of components from our analysis remains consistent with the value-added catalogue. For the negative binomial distribution, all sources have at least one component, consistent with the completeness considerations presented in Sect.~\ref{sec:DifferentialCounts}.

Interestingly, the preference for a negative binomial distribution is consistent with previous studies of counts-in-cells for optical sources (e.g., \citealt{NeymanShane1953} in the Lick survey; \citealt{SaslawYang2011} and \citealt{Hurtado-Gil2017} in the SDSS survey). \citet{Carruthers1983} proposed a universal mechanism that might link galaxy clusters counting distributions to empirical data from high-energy collisions in particle physics. While some authors (\citealt{FangSaslaw1996}) have rejected the negative binomial distribution as a physically complete description of galaxy clustering, arguing that it violates the second law of thermodynamics, others have found it to be justified (\citealt{Elizalde1992}; \citealt{Betancort-Rijo2000}; \citealt{Hurtado-Gil2017}). 

In the contex of halo occupation distribution models (HOD; \citealp{HOD2002}; \citealp{HOD2005}), which study the relationship between the number of galaxies and the mass of their host halos, these models provide a statistical framework to describe how galaxies trace dark matter halos. The HOD framework is particularly important for linking observations of galaxy clustering to the underlying dark matter distribution. Simulation results indicate that the scatter in the number of satellite galaxies at a fixed halo mass is likely non-Poissonian (\citealp{Boylan-Kolchin2010}; \citealp{Mao2015}), suggesting that simple Poisson-based assumptions may not fully capture the complexity of galaxy-halo relationships. \citet{Jimenez2019} demonstrated that a negative binomial distribution provides a better model for the number of satellite galaxies in halos. The negative binomial distribution accounts for this increased variability by introducing a second parameter that allows for more flexibility in the variance. Unlike the Poisson distribution, which assumes fixed variance, the negative binomial can handle overdispersed data, where the variance exceeds the mean.

\subsection{Comparison of the methods for two-point correlation function}\label{MethodsComparisonTPCF}

In Sect. \ref{sec: AngularCorrelationFunction}, we employed the scaling of cell sizes to determine the two-point correlation function using the reduced normalised variance. This approach simplifies the computation by using source counts within cells rather than direct pairwise comparisons. Compared to the direct measurement from the Landy-Szalay estimator used by \cite{Hale2023}, our method is computationally more efficient, scaling linearly with the number of sources, \(N\), instead of the quadratic scaling $N(N-1)/2$ required for direct pair counting. As future surveys detect increasingly larger numbers of radio sources, calculating the two-point correlation function will become significantly more time-consuming, even with advanced methods developed to address this challenge.

For comparing different bias models within the HOD framework, the method outlined in \citealp{Hale2023} (see Sects. 5 and 6 of \citealp{Hale2023}) may prove more advantageous than our approach, particularly because it employs the full estimator, which provides a more comprehensive analysis of clustering behavior. While our approach is well-suited for fitting a power law, its applicability is limited providing values for the two-point correlation function across all angular separations, as it depends on a predefined set of cell sizes, which constrain its flexibility and broader applicability. In contrast, the estimator approaches used in direct measurements are more complicated when calculating higher-order correlation functions, which are necessary for studying non-Gaussianity, especially at large scales. However, our method can be more easily extended to higher-order correlation functions, as it primarily relies on statistical moments.

\section{Conclusion} \label{sec: conclution}

In this work, we studied the counts-in-cells distribution of LoTSS-DR2 radio sources, to determine the best-fitting statistical model. By applying spatial masks and flux density thresholds, we conducted a comparative analysis of the distribution of sources using three discrete stochastic processes. We found, with high statistical significance, that the distribution of the radio sources deviates from a Poisson distribution, indicating the presence of clustering. Our research indicates that a Cox processes, specifically compound Poisson and negative binomial distributions, provide a good fit to the source distribution. To distinguish between these models, we performed two
hypothesis tests (the reduced chi-square and Kolmogorov-Smirnov tests) and calculated the 
Bayes factor. 
All three tests culminate in the conclusion that there is strong statistical evidence in favour of the negative binomial distribution for the counts-in-cells of radio sources above flux density thresholds of 2 mJy. We focused our analysis on the Poisson, compound Poisson, and negative binomial distributions, leaving open the possibility that other distributions might also fit the data well. 

By analysing the scaling properties of cell sizes and relationship between the second moment of the counts-in-cells and the two-point correlation function, we performed a fitting procedure on the reduced normalised variance across different cell sizes. This fitting process allows us to determine the parameters of the two-point correlation function, specifically identifying the amplitude and exponent of its power law. At a flux density threshold of 2 mJy and a SNR of 7.5, we measure a value of $1 - \gamma$ in the range of $-1.05$ to $-0.8$, which aligns well with previous studies using optical, infrared and radio data (\citealt{SaslawYang2011}; 
\citealp{Labini_2011};
\citealt{TaylorInfrared2004}; \citealt{Pollo2012};  \citealt{Magliocchetti1999}; \citealt{Lindsay2014}).

Since the counts-in-cells statistics clearly indicate clustering, higher-order moments of the galaxy distribution can be used to test non-linear and non-Gaussian models of the large-scale structure formation. While this work focuses on the second moment, which relates to the two-point correlation function, a more accurate description of the radio galaxy distribution requires measuring higher-order correlation functions using higher moments. Future work should address this, as pioneered in \citealt{Peebles1980}, \citealt{Magliocchetti1999} and \citealt{saslaw2000distribution}.

Applying the same method used in this work to higher-order correlation functions will scale the computational effort linearly with the number of sources, \(N\). In contrast, methods that rely on optimal estimators, such as the Landy-Szalay estimator for the two-point correlation, scale with \(N^m\) for the \(m\)-point correlation function, making them computationally expensive for future surveys with a very large number of sources. Therefore, we conclude that simple counts-in-cells methods remain valuable, as they provide resource efficient and powerful insights into the large-scale structure of the Universe. The upcoming LoTSS Data Release 3 will present an opportunity to investigate non-Gaussianity on the large cosmological scales (see e.g. \citealp{Desjacques2010}).

\begin{acknowledgements}
MPA acknowledges support from the Bundesministerium für Bildung und Forschung (BMBF) ErUM-IFT 05D23PB1. 
LB acknowledges support by the Studienstiftung des deutschen Volkes.
TMS and DJS acknowledge the Research Training Group 1620 `Models of Gravity', supported by Deutsche Forschungsgemeinschaft (DFG) and by the German Federal Ministry for Science and Research BMBF-Verbundforschungsprojekt D-LOFAR IV (grant number 05A17PBA). CLH acknowledges support from the Leverhulme Trust through an Early Career Research Fellowship and also acknowledge support from the Oxford Hintze Centre for Astrophysical Surveys which is funded through generous support from the Hintze Family Charitable Foundation. CH’s work is funded by the Volkswagen Foundation. CH acknowledges additional support from the
Deutsche Forschungsgemeinschaft (DFG, German Research Foundation) under Germany’s Excellence Strategy EXC 2181/1 - 390900948 (the Heidelberg STRUCTURES Excellence Cluster). 
JZ acknowledges support by the project "NRW-Cluster for data intensive radio astronomy:
Big Bang to Big Data (B3D) “funded through the programme "Profilbildung 2020", an
initiative of the Ministry of Culture and Science of the State of North Rhine-Westphalia.“ 

LOFAR is the Low Frequency Array designed and constructed by ASTRON. It has observing, data processing, and data storage facilities in several countries, which are owned by various parties (each with their own funding sources), and which are collectively operated by the ILT foundation under a joint scientific policy. The ILT resources have benefited from the following recent major funding sources: CNRS-INSU, Observatoire de Paris and Université d’Orléans, France; BMBF, MIWF-NRW, MPG, Germany; Science Foundation Ireland (SFI), Department of Business, Enterprise and Innovation (DBEI), Ireland; NWO, The Netherlands; The Science and Technology Facilities Council, UK; Ministry of Science and Higher Education, Poland; The Istituto Nazionale di Astrofisica (INAF), Italy.

This research was conducted using several tools and Python packages that were crucial for our analysis, including:
healpy \citep{Healpy},
HEALPix \citep{Healpix},
Astropy (\citealt{Astropy}, \citealt{Astropy2018}, \citealt{Astropy2022}), NumPy \citep{Numpy},
SciPy \citep{Scipy}, TreeCorr \citep{Jarvis2015},
IPython \citep{iPython:2007},  
Matplotlib \citep{Matplotlib}. We also used Aladin \citep{Aladin2000} to produce some images.

\end{acknowledgements}

\bibliographystyle{aa} 
\bibliography{aa} 

\appendix

\section{Masking strategies}\label{sec:Masks}

\begin{figure*}
\includegraphics[width=\linewidth]{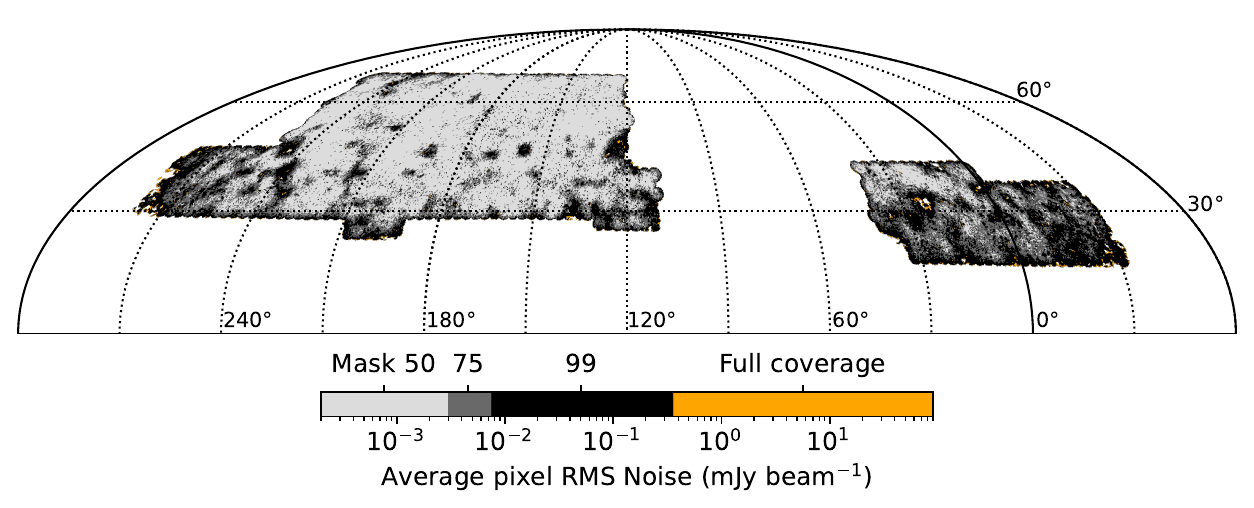}
\caption{Noise masks for LoTSS-DR2. The grey regions correspond to the lowest noise regions included in `mask50', while dark grey is included in `mask75' and black in `mask99'. The orange regions are outside of all three masks but non-empty cells in LoTSS-DR2.}
\label{fig:noise}
\end{figure*}

\begin{table*}
    \centering
    \caption{Comparison of the different masks for the LoTSS-DR2. The full sky with a {\sc HEALPix} resolution of $N_\mathrm{side}=256$ has $786\,432$ cells.}
    \begin{tabular}{ccccc}
    \hline\hline
    \addlinespace[0.5ex]    Mask  &  $N_\mathrm{cell}$   &   $f_\mathrm{sky}$   &   $N$  &   Description\\
            &   & \% & &      \\\addlinespace[0.5ex]\hline
        \addlinespace[0.5ex]MOC &   $107\,814$  & $13.7$ & $4\,370\,692$ & Coverage defined by MOC \\\addlinespace[0.5ex]\hline
        \addlinespace[0.5ex]5s   &   $105\,958$  &  $13.5$& $4\,367\,156$   & MOC \& Exclude cells with less than five sources \& App.~\ref{sec:threecells}  \\\addlinespace[0.5ex]\hline
        \addlinespace[0.5ex]99  &   $104\,898$  &$13.3$ & $4\,353\,229$   &   Mask 5s \& Exclude cells with $S_\mathrm{rms}>0.27$~mJy beam$^{-1}$  \\\addlinespace[0.5ex]
        \addlinespace[0.5ex]75  &   $79\,468$   & $10.1$& $3\,768\,962$   & Mask 5s \&  Exclude cells with $S_\mathrm{rms}>0.11$~mJy beam$^{-1}$\\\addlinespace[0.5ex]
        \addlinespace[0.5ex]50  &   $52\,979$   & $6.7$& $2\,868\,431$   & Mask 5s \&  Exclude cells with $S_\mathrm{rms}>0.07$~mJy beam$^{-1}$\\\addlinespace[0.5ex]\hline
        \addlinespace[0.5ex] 
        d  & $84\,625$   & $10.7$ & $3\,685\,561$  & Mask 5s \& Exclude area outside of outer pointing centres
        \\\addlinespace[0.5ex]\hline\addlinespace[0.5ex]
        1219  &  $98\,746$   & $12.5$ & $4\,176\,162$   &  Mask 5s \& Include cells with $\theta_1 < 1.2$ deg, $\theta_2 < 1.9$ deg \\
        \addlinespace[0.5ex]\addlinespace[0.5ex]
        1017  &  $82\,011$   & $10.4$ & $3\;494\;360$   &  Mask 5s \& Include cells with $\theta_1 < 1.0$ deg, $\theta_2 < 1.7$ deg \\\addlinespace[0.5ex]\hline
     
    \end{tabular}
    \label{tab:masks}
\end{table*}

The sky coverage, observed by the LoTSS-DR2, is available as a multi-order coverage map (MOC; \citealt{MOC2014}).
We use this MOC to define a {\sc HEALPix} mask with a resolution of $N_\mathrm{side}=256$, which corresponds to a covered area of $\Omega_{\mathrm{cell}}\approx1.60\times10^{-5}$~sr $\approx 189.1$ square arcmin for each cell.
Initially, the highest resolution of the MOC is $N_\mathrm{side}=1024$. We then downgrade the resolution to $N_\mathrm{side}=256$ and reject all cells in the new resolution, which already have one masked cell in the old resolution.
This procedure ensures that the new coverage mask only recovers observed sky patches, but also decrease the usable sky fraction.
Fig.~\ref{fig:fullDR2} shows the LoTSS-DR2 radio source catalogue, displayed as a {\sc HEALPix} source count map, without applying any mask or flux density threshold.
With the MOC mask, we recover $4\,370\,692$ sources within a fractional sky coverage of $f_\mathrm{sky}\approx13.7\%$, which corresponds to $\sim1.72$~sr.
In addition to the coverage defined by the MOC mask, we exclude cells with less than five sources in it to ensure statistical stability and reject (nonphysical) empty cells (see \citealt{Siewert2019} for more details). This mask will be called `mask 5s'.

\subsection{Default mask}

The LoTSS-DR2 source catalogue is based on a mosaic of many individual pointings. Typically maximal sensitivity is achieved close to the pointing centre. Combining the information from individual overlapping pointings reduces the effects of decreasing sensitivity and beam smearing in the outer regions of the individual observations of a pointing. In the mask d (see Fig.~\ref{fig:boundary_poinitings} and Tab.~\ref{tab:regions}) we remove area close to the boundary of the survey area in which the mosaic cannot use information form more than a single pointing. More details of the mask d are described in 
Sect.~\ref{sec:masking_strategies} and in \citet{Hale2023}.

\subsection{Masks based on rms noise}\label{sec:rmsmasks}
Similar to the set of masks of \citet{Siewert2019}, we define a set of masks for the full LoTSS-DR2 based on the local rms noise of the sources averaged per cell.
We define three noise masks that keep 50\%, 75\%, and 99\% of all cells ranked by increasing averaged cell noise.  
The masks are named `mask 50', `mask 75', and `mask 99', respectively.
Additionally, we reject in all noise masks three cells with a large amount of components arising from large and nearby spiral galaxies and clusters. For more detail on these three cells, see App.~\ref{sec:threecells}.
The five masks are presented in Fig.~\ref{fig:noise} and the details of the observed number of cells ($N_\mathrm{cell}$) and sources ($N$) for the LoTSS-DR2 are presented in Table~\ref{tab:masks}.
The low noise region of `mask 50' with an averaged local noise of $S_\mathrm{rms}\leq 0.086$~mJy per cell corresponds to the grey region in Fig.~\ref{fig:noise}. 
Including medium grey and black regions correspond to `mask 75' and `mask 99' with $S_\mathrm{rms}\leq0.111$~mJy and $0.274$~mJy, respectively. Including the dark blue regions defines the full coverage of the LoTSS-DR2.

For the comparison of LoTSS-DR1 and DR2, we make use of the masks defined in \citet{Siewert2019}, see Fig.~\ref{fig:ComparisonHetdex}.
As the first data release of the LoTSS radio source catalogue has an incomplete coverage in the HETDEX spring field and under-sampled regions, we have to mask those regions by rejecting cells of the DR2 source count map.
In order to compare consistent sky coverages, we apply the default mask of the cosmology analysis of LoTSS-DR1 (`mask d'; \citealt{Siewert2019}) also to the LoTSS-DR2.
This mask rejects five incomplete and under-sampled pointings, based on an averaged pointing radius of $r=1.7$~deg, and, additionally, rejecting cells with less than five sources.
The sky coverage of this mask is $f_\mathrm{sky}=0.9\%$, which corresponds to $7176$ observed cells.

\subsection{Masks based on pointing geometry}\label{sec:pointingGeometry}

Our third masking strategy is based on the geometry of the pointings and the distance of the {\sc HEALPix} cell centres from the pointing centres. 
As discussed before the variations in flux scale across the surveyed area are reduced by mosaicing. 
It means that overlapping reduces the variations of the flux scale. 
Based on it we define our strategy as follow. Below a given angular distance between the pointing and cell centre ($\theta_1$) we accept all cells. 
For cells further away, but below a second angular distance ($\theta_2$), we accept also all cells which are the neighbourhood of at least two pointing centres, which means overlapping of at least two pointings. 
Additionally, following \citet{Siewert2019}, we exclude cells with less than five sources in them to reject unexpectedly empty cells and we reject the three cells described in the App.~\ref{sec:threecells}.

\begin{figure}
    \centering
    \includegraphics[width=\linewidth]{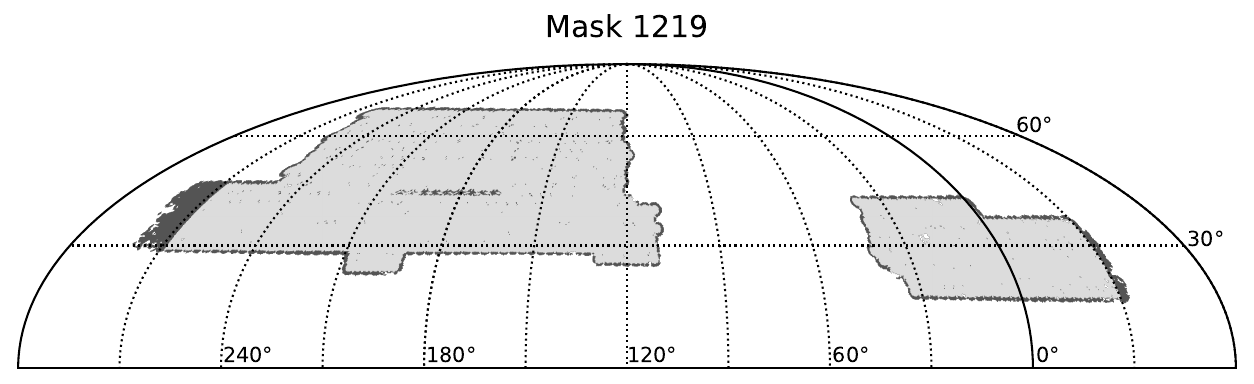}
    \includegraphics[width=\linewidth]{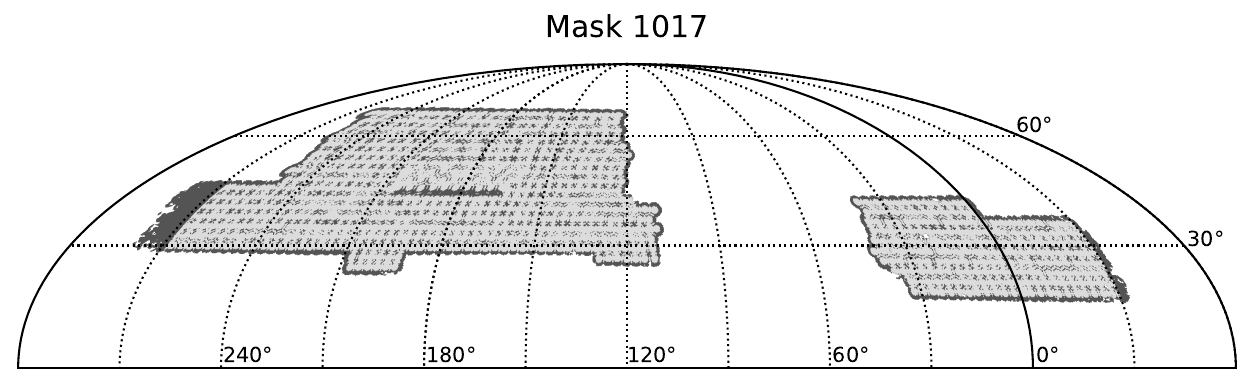}
    \caption{The 'mask 1219' (\textit{top}) and 'mask 1017' (\textit{bottom}) are based on the survey and pointing geometry. The light grey regions remain after masking. The patterns shown in the bottom panel represent masked regions that lie far from the pointing centre and are not shared by at least two pointings.}
    \label{fig:enter-label}
\end{figure}
For this pointing based masks we use two sets of combinations, which are shown in Fig.~\ref{fig:1017_1219masks}.
A more conservative one which leads to a quite patchy accepted survey area and one that minimises the number of holes in the survey area.
Again, we made use of the {\sc HEALPix} pixelation in $N_\mathrm{side} = 256$. 
For the conservative mask, we take $\theta_1 = 1$~deg and $\theta_2 = 1.7$~deg. 
It means that until 1~deg from pointing centres we pick all cells and from 1 to 1.7~deg we pick just the cells which are in common in at least two pointings. 
We call it `mask 1017', which keeps $76\%$ of the survey area. The number 10 in the mask name corresponds to 1.0 deg., and 17 corresponds to 1.7 deg. representing $\theta_1$ and $\theta_2$ in the mask definition. For the less conservative mask we have $\theta_1 = 1.2$~deg and $\theta_2 = 1.9$~deg. This mask is called `mask 1219' and it keeps $92\%$ of the survey area. As before, the number 12 in the mask name corresponds to $1.2$~deg. and 19 corresponds to $1.9$~deg., representing $\theta_1$ and $\theta_2$ in the mask definition. Both of these two combinations maintain an additional cut at $\mathrm{RA}=270$~deg for the RA-13~h field from the former mask, as the pointings outside this cut are mainly incomplete, which can be seen in Fig.~\ref{fig:boundary_poinitings} and \ref{fig:1017_1219masks}.

\section{Excluded Cells}\label{sec:threecells}

\begin{figure}
    \centerline{\includegraphics[width=0.8\linewidth]{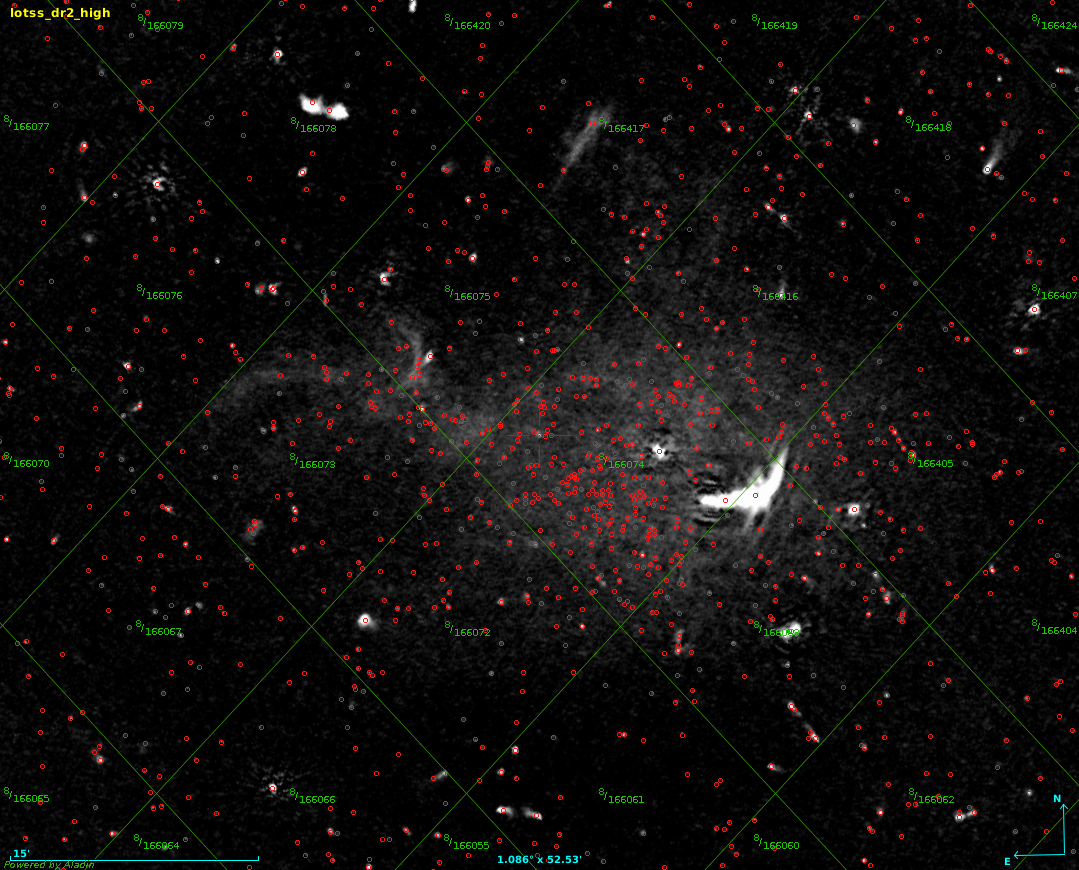}}
    \centerline{\includegraphics[width=0.8\linewidth]{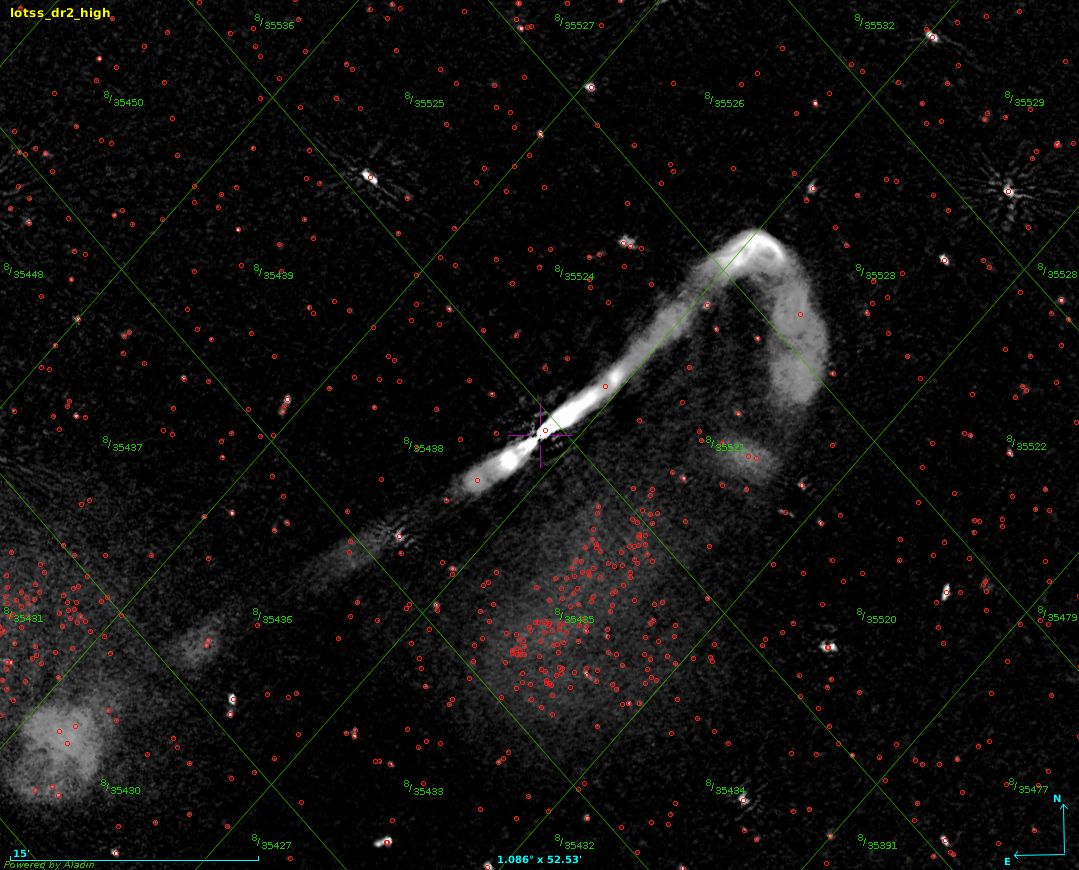}}
    \centerline{\includegraphics[width=0.8\linewidth]{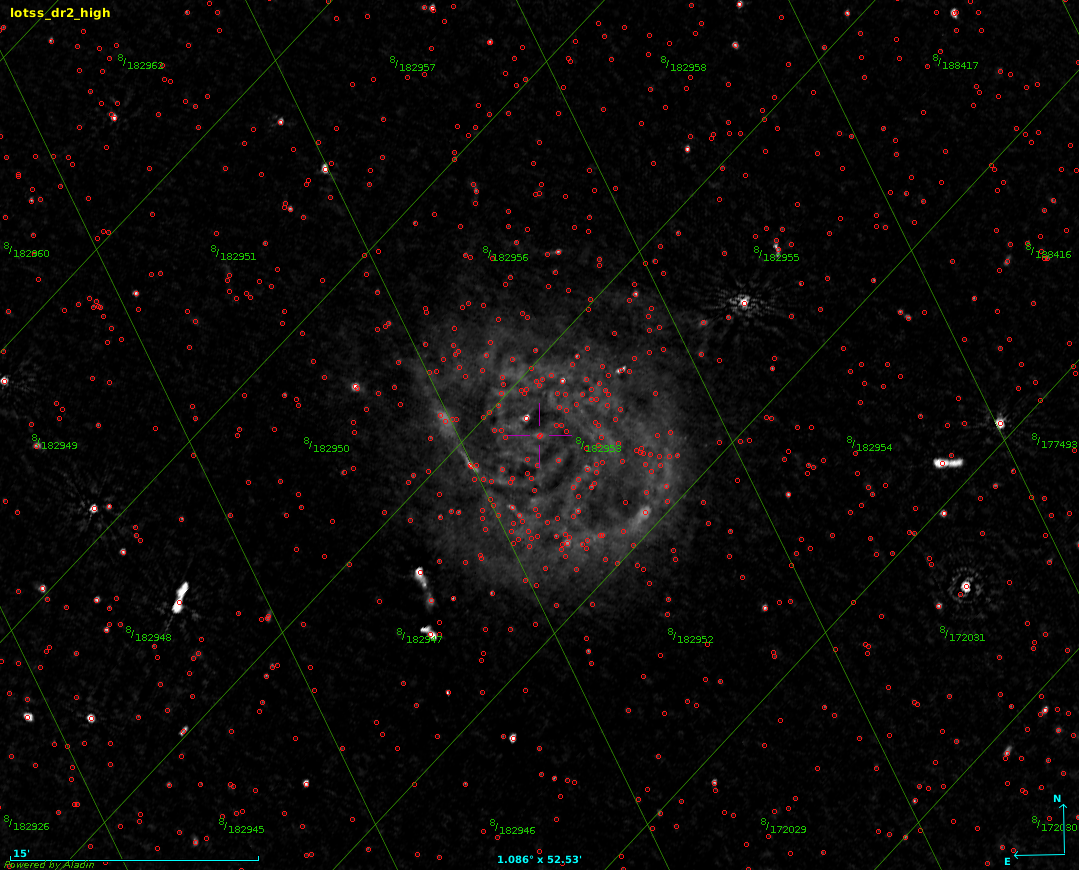}}
    \caption{NGC4889 galaxy in Coma cluster with diffuse emission around it \textit{(top)}, NGC315 galaxy, an active AGN with its jets \textit{(middle)}, NGC5457 or M101 or Pinwheel galaxy, a nearby galaxy \textit{(bottom)}.
    The red circles are the sources detected by {\sc PyBDSF}. The green grid is the {\sc HEALPix} cells in $N_\mathrm{side}=256$.}
    \label{fig:three_largest_pixels}
\end{figure}

The cell with the highest source density is cell $208\,938$ in {\sc HEALPix} $N_\mathrm{side}=256$ ring scheme with 231 sources. 
It is located in the pointing P195+27 and contains the diffuse emission near the galaxy NGC 4889, which is deeply embedded within the Coma cluster. 
NGC 4889 is a super-giant elliptical galaxy well-known since the 18th century. 
Fig.~\ref{fig:three_largest_pixels} top panel shows NGC 4889 with a red cross detected by the Python Blob Detection and Source Finder algorithm \citep[\sc PyBDSF]{pybdsf} as separated sources. 
The small red dots are the detected sources and the green grid shows the {\sc HEALPix} cells in $N_\mathrm{side}=256$ in the nested scheme. 
The figure is generated by {\sc Aladin\footnote{ {https://aladin.u-strasbg.fr/}}}. {\sc Aladin} uses the nested scheme of {\sc HEALPix}, this is why the labels are different.

The second highest source density is found in cell $195\,113$ with $155$ sources. 
In this case the source finder picked up the emission of jet gas of an AGN, 
the galaxy NGC 315 (Fig.~\ref{fig:three_largest_pixels} middle panel). 
It is a giant elliptical galaxy, also well-known since the 18th century and was discovered by William Herschel. 
Its distance is $\sim 70~\mathrm{Mpc}$ \citep{NGC315}. 

The third cell that we mask is $73\,793$ in pointing P209+55 with 141 sources. 
Here the source finder picks up emission from gas and stars of the well-known Pinwheel galaxy, NGC5457 or M101 (Fig.~\ref{fig:three_largest_pixels} bottom panel), 
a nearby galaxy of a distance of $6.4~\mathrm{Mpc}$ \citep{NGC5457}. {\sc PyBDSF} does not detect it as a single source but many sources. 

We checked that the fourth and fifth most populated cells sit in the smoothly extrapolated tail of the empirical counts-in-cells distribution and are kept for that reason. For calculation of the one- and two-point functions we should therefore exclude the three most populated cells.

 \section{Comparison to DR1}\label{app:DR1}
\begin{figure}
    \centering
    \includegraphics[width=\linewidth]{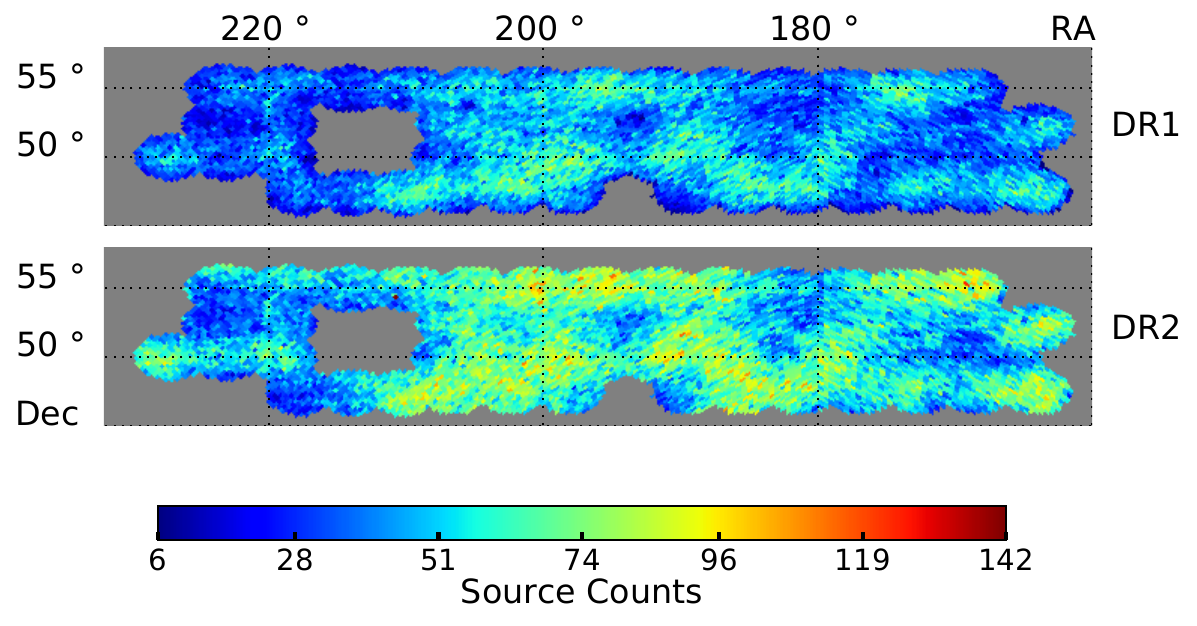}
    \caption{Source counts per {\sc HEALPix} cell of the LoTSS-DR1 (\textit{top}) and LoTSS-DR2 (\textit{bottom}) for the HETDEX spring field. Both surveys are masked with the default mask of \citet{Siewert2019}. The observed difference in the mean source counts is attributed to changes in the pipeline for DR2 \citep{Shimwell2022}.} 
    \label{fig:ComparisonHetdex}
\end{figure}
 \subsection{Comparison to HETDEX field}

Similar to \cite{Siewert2019}, we use the empirical moments to examine the region that was investigated in LoTSS-DR1 to identify the effect that the improved analysis pipeline has on the counts-in-cells statistics. For details on the definition of the sample statistics see Sect.~\ref{sec:statmoments} and \ref{sec:cic_dist}.

\begin{figure*}
    \centering
    \includegraphics[width=0.9\linewidth]{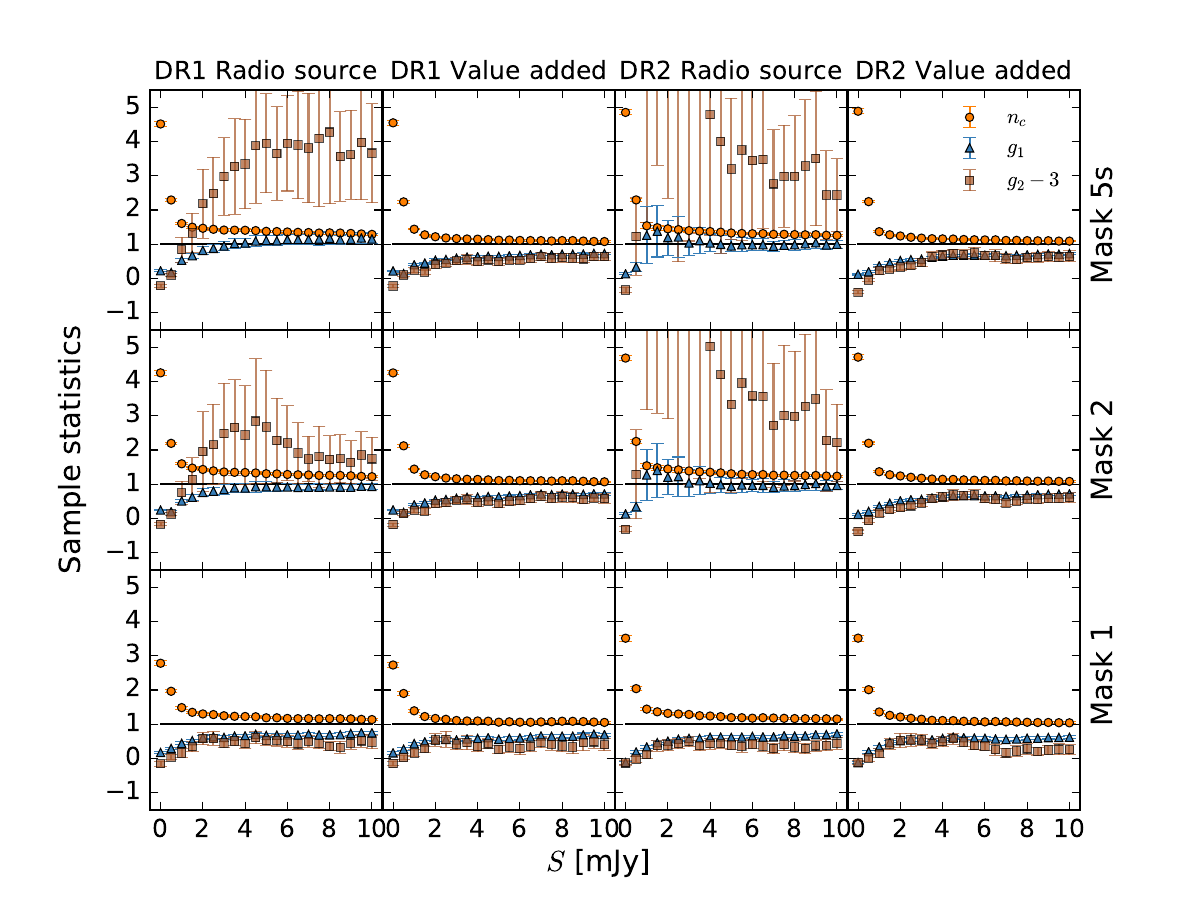}
    \caption{Empirical moments of the LoTSS-DR1 radio source catalogue (\textit{left}), DR1 value-added catalogue (\textit{middle left}), the LoTSS-DR2 radio source catalogue (\textit{middle right}), and DR2 value-added catalogue (\textit{right}). We show the clustering parameter ($n_c$, orange), skewness ($g_1$, blue), and excess kurtosis ($g_2-3$, brown) for the HETDEX field of DR1 and DR2 with `mask 1', `mask 2', and `mask 5s' of \citet{Siewert2019}.
    }
    \label{fig:empiricalmomentsHETDEX}
\end{figure*}

The distribution of source counts in LoTSS-DR1 revealed that $n_c > 1$, indicating an excess in variance compared to a Poisson distribution for all flux density thresholds considered. This showed good agreement with a compound Poisson distribution \citep{Siewert2019}. While the improvements of the second data release lead to an increased source density in the HETDEX field, we calculate the empirical moments, such as clustering coefficient, skewness, and excess kurtosis of DR2 and compare them to the results of DR1 \citep{Siewert2019}. In Fig.~\ref{fig:ComparisonHetdex}, we show the source count map of the HETDEX field covered by 'mask 5s' of \citet{Siewert2019} (originally named 'mask d') for both data releases.
The sample statistics of both radio source catalogues as function of flux density threshold are compared in Fig.~\ref{fig:empiricalmomentsHETDEX}.

Additionally, we present the sample statistics for both the DR1 value-added catalogue and the DR2 value-added catalogue. It is important to note that the DR2 value-added catalogue only looks to the sources brighter than 8 mJy, rather than all sources in the DR2 catalogue.
In order to compare the sample statistics of the same regions in the HETDEX field, 
we also use the noise masks of \citet{Siewert2019}, which reject cells with a per cell averaged local noise above one (`mask 1') or two times the median local noise (`mask 2') of $71~\mu$Jy~beam$^{-1}$.
As already discussed in \cite{Siewert2019}, the excess kurtosis (brown boxes) of the DR1 sample statistics depends on the local noise of the field of view and depends on improvements made in the value-added catalogue using a combination of statistical techniques and visual association and identification. Thereby reducing outliers in the distribution and resulting in a smoother kurtosis. 
Also the clustering coefficient (orange points) comes closer to the expectation for a Poisson distribution of $n_c=1$ for the value-added catalogue and the lowest noise region. 
For the HETDEX field of the DR2 radio source catalogue, we find a similar difference between `mask 5s' and `mask 1'. 
The intermediate cut in local noise of `mask 2' still shows a highly varying excess kurtosis and skewness (blue triangles).
\begin{figure}
    \centering
    \includegraphics[width=\linewidth]{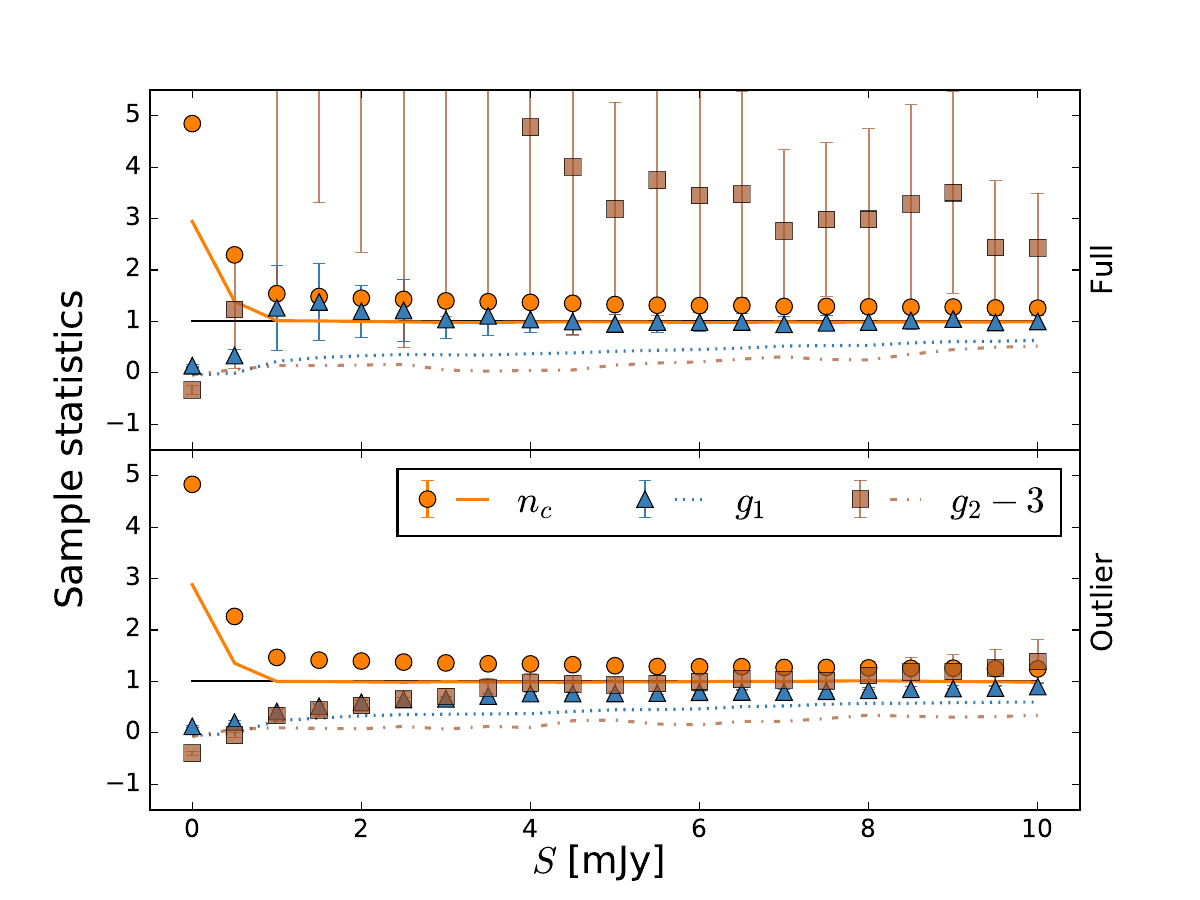}
    \caption{Empirical moments of the LoTSS-DR2 radio source catalogue (\textit{top}) for the HETDEX field and the same catalogue, but with the most dense cell removed (\textit{bottom}), as described in Sect.~\ref{sec:rmsmasks} and App.~\ref{sec:threecells}. Lines represent the corresponding sample statistics of a mock sample of same size.}
    \label{fig:empiricalmomentsOutlier}
\end{figure}

The most prominent effect to the sample statistics can be seen by removing one cell with the highest source density. 
This cell, with the index `73793' in the {\sc HEALPix} ring scheme, covers the Messier 101, or NGC5457, spiral galaxy. 
For more details see App.~\ref{sec:threecells}. 
By removing this cell, the sample statistics, especially the excess kurtosis, become more continuous over all flux density thresholds. 
In Fig.~\ref{fig:empiricalmomentsOutlier} this effect is shown for `mask 5s' of \citet{Siewert2019} for the complete DR2 HETDEX sample (top) and for the sample with the cell removed (bottom). The sample statistics of a randomly drawn sample of same size from a mock catalogue (see Sect.~\ref{sec:mock}) are also shown as lines. Large multi-component sources, which increase the cell source density in the catalogue, are not simulated in the mock catalogue. Therefore the mock catalogue is stable against removing the same most dense cell of the catalogue, like seen in Fig.~\ref{fig:empiricalmomentsOutlier}. In general the clustering parameter of the mock sample represents for most flux density thresholds a Poisson distribution with $n_c=1$. At flux density thresholds below 1~mJy, the variance of the counts-in-cell distribution of the mock catalogues increases, which corresponds to the increasing incompleteness at lower flux densities. 

\subsection{Empirical moments of the full LoTSS-DR2}

\begin{figure*}
    \centering
    \includegraphics[width=0.8\linewidth]{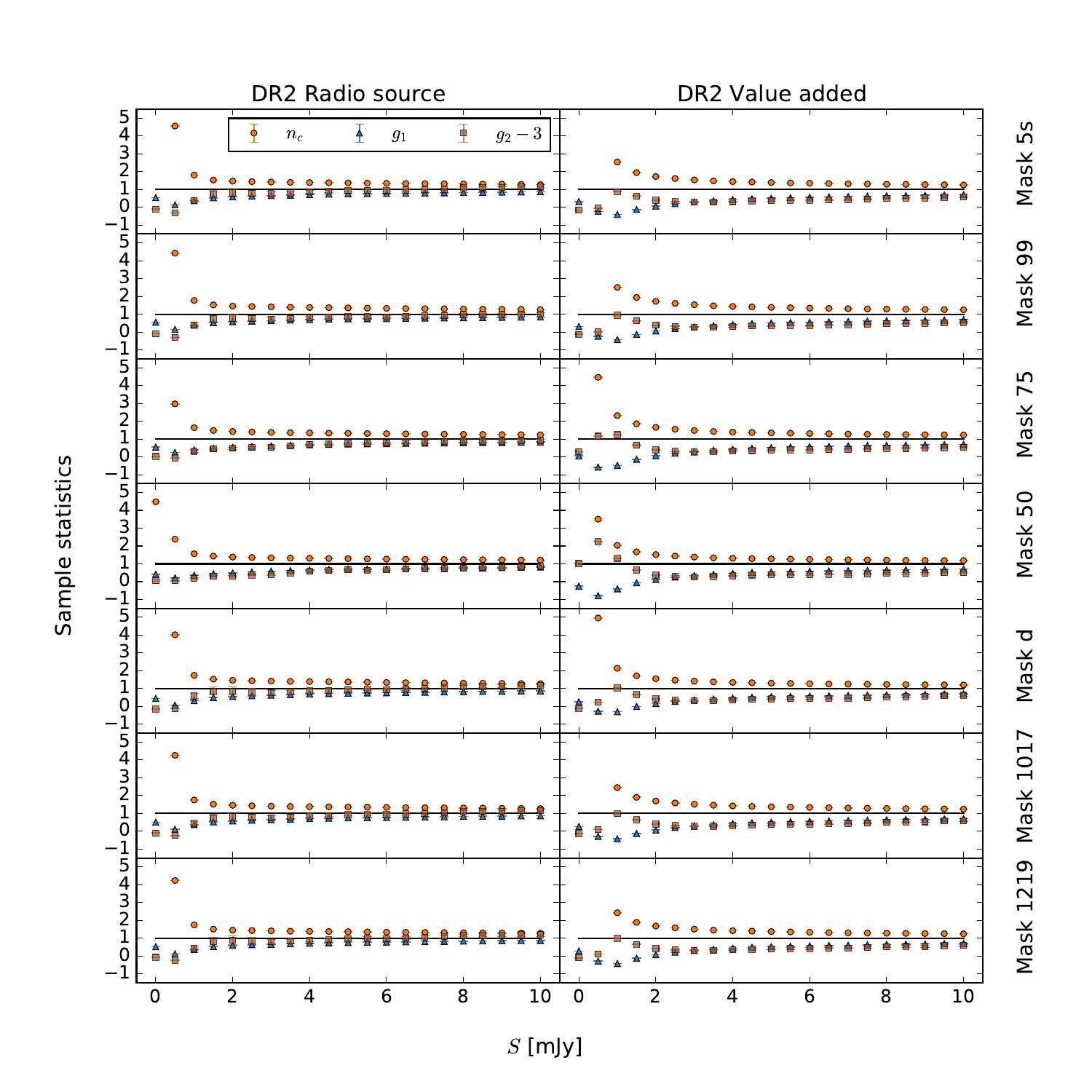}
    \caption{ Sample statistics of the full LoTSS-DR2 radio source catalogue  (\textit{left}) and value-added catalogue (\textit{right}) as function of flux density threshold for all masks defined in Table~\ref{tab:masks}. We show the clustering parameter ($n_c$, orange), skewness ($g_1$, blue), and excess kurtosis ($g_2-3$, brown).
    \label{fig:samplestatistics_full}}
\end{figure*}

In comparison to the smaller HETDEX field (see Fig.~\ref{fig:empiricalmomentsHETDEX} and Fig.~\ref{fig:empiricalmomentsOutlier}), the sample statistics of the full LoTSS-DR2 field (see Fig.~\ref{fig:samplestatistics_full}) show a different behaviour while masking regions with higher noise. The skewness and excess kurtosis are for all masks similar and more continuous for flux density thresholds between 2 and 10~mJy, than seen before in the smaller region of the HETDEX field. The clustering parameter still varies with values $n_c>1$ from the general assumption of a pure Poisson distribution, for which $n_c=1$ holds. Two effects seem reasonable for the more continuous skewness and excess kurtosis. Firstly, the noise masks additionally exclude three cells with high number of sources (see App.~\ref{sec:threecells}). Removing these outliers of the Poisson distribution can smooth the sample statistics over a broader range of flux density thresholds.
The lower statistical moments are defined on the sample mean and are therefore sensitive to outliers. Secondly, in the case of the larger sample size of the full LoTSS-DR2, the underlying distribution is sampled with more statistical accuracy for each subset of flux density thresholds. For all masks at flux densities above 2 mJy, the skewness is about +1, indicating that the distribution is slightly skewed towards higher counts-in-cells. It differs from the skewness of the Poisson distribution, whose expectation value is close to zero for the flux threshold of 2 mJy (see Eq.~\ref{eq:5} and the values at the Table~\ref{tab:Distribution_parameters}).
Similar behaviour of the clustering parameter, the excess kurtosis and the skewness can be observed for all spatial masks. The behaviour of the statistical moments of the noise masks (`mask 50', `mask 75', and `mask 99') is almost the same and stable for flux densities above 2mJy. This is another reason why we decide to use mask d for the cosmological analysis of LoTSS-DR2.

\section{Results of distribution  and clustering analysis for masks 1219 and 1017}\label{sec:Geometry-based_masks_results}

\subsection{Counts-in-cells distribution analysis}

In this section, we present the results of the counts-in-cells distribution analysis, similar to Sect.~\ref{sec:countsincellsdistribution}, focussing on masks 1219 and 1017. Table \ref{tab:Distribution_parameters_Geo_masks} lists the parameters for the Poisson, compound Poisson, and negative binomial distributions at various flux density thresholds. The reduced chi-square test results for these masks are provided in Table \ref{tab:ChiSquared_Geo_masks}. Compared to mask d (see Table \ref{tab:ChiSquared}), masks 1219 and 1017 yield reduced chi-square values closer to one. Notably, only the 2 mJy flux density threshold for these masks results in values below two. This indicates that masks 1219 and 1017 may effectively mitigate flux density fluctuations towards the outer regions of the pointings (see \citealp{Shimwell2022}), thereby reducing systematic effects.

\begin{table}
\label{tab:Distribution_parameters_Geo_masks}
	\caption{Values of the parameters of three different models for the counts-in-cells distribution of LoTSS-DR2 sources.}
	
    \setlength{\tabcolsep}{4.5pt}
    \renewcommand{\arraystretch}{1} 
    
	\begin{tabular}{ccc|cc|cc}
		\hline \hline
        Mask &
		\multicolumn{1}{p{1.5cm}}{\centering $S_\mathrm{min}$} & 
        \multicolumn{1}{p{1cm}}{\centering Poisson} &
        \multicolumn{2}{p{1cm}}{\centering compound \\ Poisson} &
        \multicolumn{2}{p{1cm}}{\centering negative \\ binomial} \\
		\cmidrule(lr){3-7}
        \addlinespace[0.5ex]
         & &
		\multicolumn{1}{p{1.3cm}}{ \centering $\lambda$} & 
        \multicolumn{1}{p{0.4cm}}{\centering $\lambda$} &
        \multicolumn{1}{p{0.4cm}}{ \centering $\kappa$} &
        \multicolumn{1}{p{0.4cm}}{\centering r} &
        \multicolumn{1}{p{0.4cm}}{\centering p}\\
		\hline
		\addlinespace[0.5ex]
		
		\multirow{4}{3em}{mask 1219} 
        & 2 mJy & $9.87$ & $21.48$ & $0.46$ & $21.48$ & $0.32$ \\ & 4 mJy & $5.51$ & $14.28$ & $0.38$ & $14.28$ & $0.28$ \\
		& 8 mJy & $3.33$ & $10.91$ & $0.31$ & $10.91$ & $0.23$\\
		\hline
		\addlinespace[0.5ex]
		\multirow{4}{3em}{mask 1017} 
        & 2 mJy & $9.87$ & $21.26$ & $0.46$ & $21.26$ & $0.32$ \\ & 4 mJy & $5.51$ & $14.05$ & $0.39$ & $14.05$ & $0.28$ \\
		& 8 mJy & $3.33$ & $10.76$ & $0.31$ & $10.76$ & $0.24$\\\hline
	\end{tabular}
\end{table}

\begin{table}
	\caption{$\chi^2$/dof for mask 1219 and mask 1017 at different flux density thresholds for LoTSS-DR2 sources. The degrees of freedom are $N_\mathrm{bins} - 1$ for the Poisson distribution and $N_\mathrm{bins} - 2$ for the compound Poisson and negative binomial distributions}.
	\begin{tabular}{cccccc}
		\hline \hline
        Mask &
		\multicolumn{1}{p{0.9cm}}{\centering $S_\mathrm{min}$} & 
        \multicolumn{1}{p{0.9cm}}{\centering $N_\mathrm{bins}$} &
        \multicolumn{1}{p{0.9cm}}{\centering Poisson} &
        \multicolumn{1}{p{1.3cm}}{\centering compound \\ Poisson} &
        \multicolumn{1}{p{1cm}}{\centering negative \\ binomial} \\
		\hline
		\addlinespace[0.5ex]
		\multirow{4}{3em}{mask\\1219} 
        & 2 mJy & $16$ & $323.7$ & $4.2$ & $1.8$ \\ & 4 mJy & $12$ & $327.1$ & $6.6$ & $2.4$ \\
		& 8 mJy & $10$ & $310.8$ & $8.6$ & $3.3$\\
		\hline
		\addlinespace[0.5ex]
		\multirow{4}{3em}{mask\\1017} 
        & 2 mJy & $16$ & $295.7$ & $4.3$ & $1.9$ \\ & 4 mJy & $12$ & $293.9$ & $7.0$ & $2.8$ \\
		& 8 mJy & $9$ & $262.6$ & $10.3$ & $4.1$ \\
		\hline		
	\end{tabular}
	\label{tab:ChiSquared_Geo_masks}
\end{table}

Following the methodology outlined in Sect. \ref{sec:countsincellsdistribution}, we apply the KS test to masks 1219 and 1017. The results are summarised in Table \ref{tab:KStest_table_Geo_masks}. Mask 1017, our most conservative mask designed to minimise imaging artefacts, and mask 1219, slightly less conservative but more restrictive than mask d, yield $d_n$-values that are notably close to the critical thresholds. This suggests that counts-in-cells filtered by these geometric masks are better described by the negative binomial distribution than by any other model considered in this study.
 
\begin{table*}
\centering	\caption{Kolmogorov-Smirnov test statistic for masks 1219 and 1017 at different flux density thresholds of the LoTSS-DR2 sources. $d_{n}$ denotes the measured test statistic and $d_{\alpha}$ its critical value at 99\%  confidence level. 
 }
	
	\begin{tabular}{cccc|cc|cc}
      \hline \hline
        Mask &
		\multicolumn{1}{p{1.5cm}}{\centering $S_\mathrm{min}$} & 
        \multicolumn{2}{p{2.5cm}}{\centering Poisson} &
        \multicolumn{2}{p{2.cm}}{\centering compound \\ Poisson} &
        \multicolumn{2}{p{2cm}}{\centering negative \\ binomial} \\
		\cmidrule(lr){3-8}
        \addlinespace[0.5ex]
         & &
		\multicolumn{1}{p{1.3cm}}{ \centering $d_n$} & 
        \multicolumn{1}{p{1.cm}}{\centering $d_{\alpha}$} &
        \multicolumn{1}{p{1cm}}{ \centering $d_n$} &
        \multicolumn{1}{p{1cm}}{\centering $d_{\alpha}$} &
        \multicolumn{1}{p{1cm}}{\centering $d_n$} & 
        \multicolumn{1}{p{1.cm}}{ \centering $d_{\alpha}$}\\

		\hline
		\addlinespace[0.5ex]
		\multirow{4}{3em}{mask 1219} 
        & 2 mJy & $0.0510$ & $0.0029$ & $0.0047$ & $0.0027$ & $0.0029$ &  $0.0027$\\ 
        & 4 mJy &$0.0461$ & $0.0028$ & $0.0062$ &  $0.0025$ & $0.0037$ & $0.0027$ \\
		& 8 mJy &$0.0379$ &  $0.0029$ & $0.0041$ & $0.0023$ &  $0.0029$ & $0.0027$\\
		\hline
		\addlinespace[0.5ex]
		\multirow{4}{3em}{mask 1017} 
        & 2 mJy & $0.0520$ & $0.0031$ & $0.0053$ & $0.0028$ & $0.0028$ & $0.0029$\\ 
        & 4 mJy &$0.0469$ & $0.0030$&  $0.0064$ &  $0.0027$ & $0.0039$  & $0.0028$\\
		& 8 mJy & $0.0380$ & $0.0029$ & $0.0046$ &  $0.0026$ & $0.0028$ & $0.0029$\\\hline
	\end{tabular}
	\label{tab:KStest_table_Geo_masks}
\end{table*}

\subsection{Clustering analysis}

\begin{figure}
    \includegraphics[width=\linewidth]{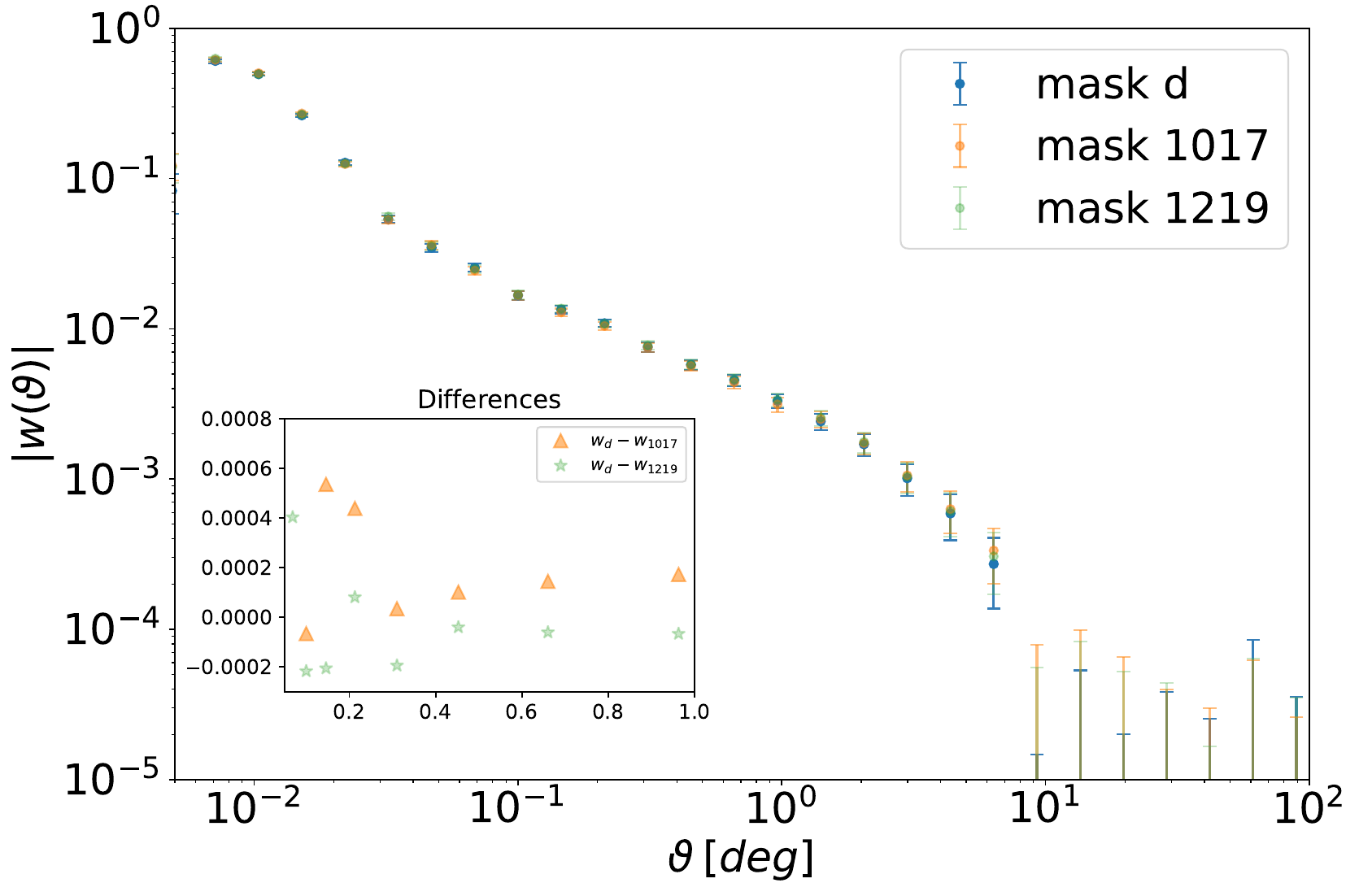}
    \caption{Comparison of the two-point correlation function for mask d, mask 1017 and mask 1219 using the Landy-Szalay estimator.}
    \label{fig:TPCF_3masks}
    
\end{figure}

In this section, we present the results of the two-point correlation function for three different spatial masks for comparison. As shown in Fig. \ref{fig:TPCF_3masks}, the results were obtained using the {\sc TreeCorr} package with the same parameter settings described in Sect. \ref{sec: AngularCorrelationFunction}. The findings indicate no significant differences between the masks, with deviations at the level of 0.01 sigma, further supporting our choice of mask d as the default for our cosmological analysis.   The inner plot highlights the differences between mask d and masks 1017 and 1219, respectively, in the range of 0.1 to 1 degree.

\section{Numerical calculation of $C_\gamma$}\label{sec:cgamma}

\begin{figure}
    \includegraphics[width=0.94\linewidth]{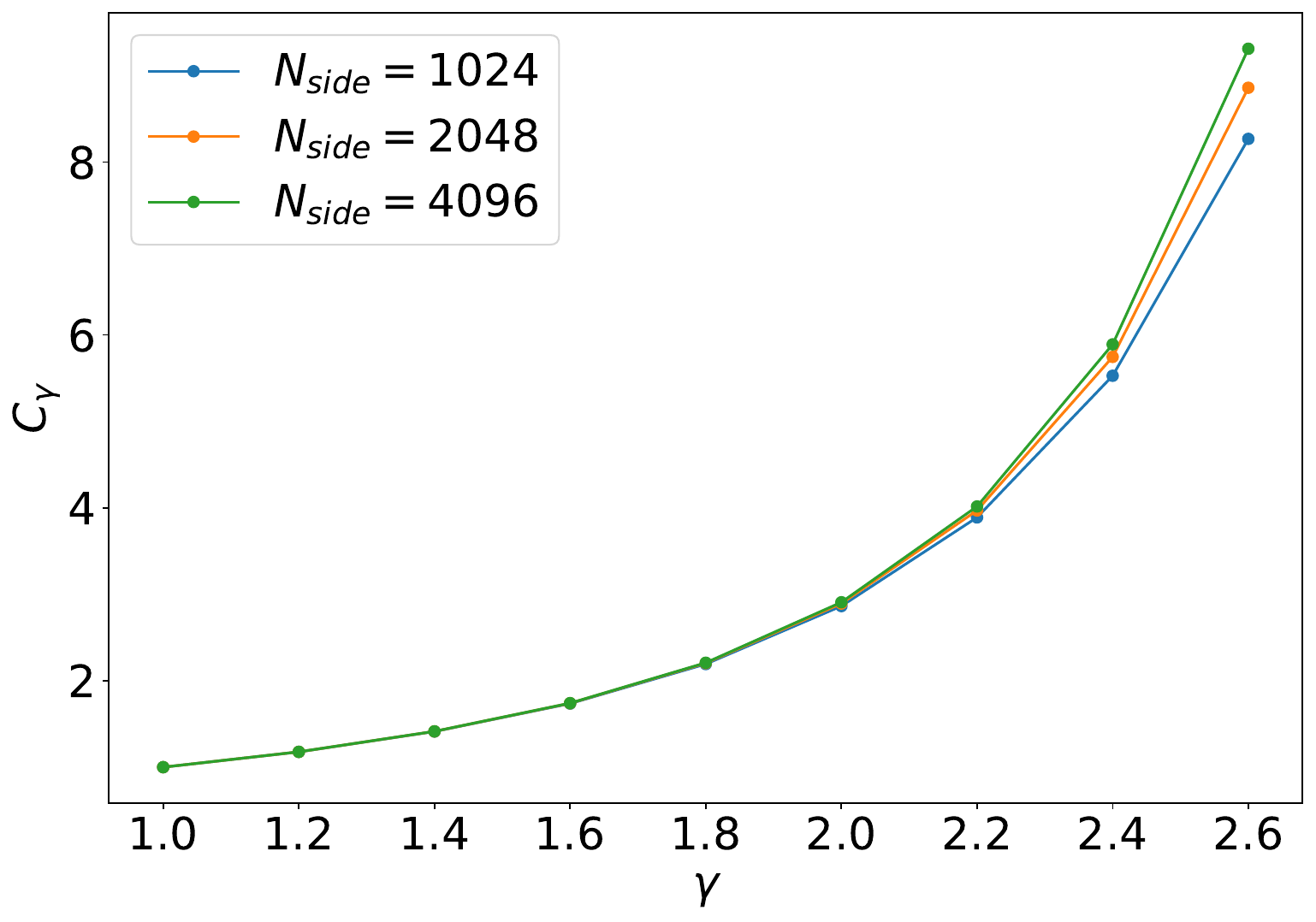}
    \caption{$C_{\gamma}$ as a function of $\gamma$. The cell at $N_\mathrm{side}=16$ is split into three different smaller cell sizes to test the variation of $C_{\gamma}$ w.r.t. $\gamma$. }
    \label{fig:Cgamma}
\end{figure}

We calculate the numerical coefficient $C_\gamma$ replacing the integral in Eq.~\ref{eq:23} by a sum over small and equally sized pixels, which provides the relation
\begin{equation}
     C_\gamma = \frac{1}{\Omega_c^2} \left(\frac{\Theta_0}{\Theta}\right)^{1 - \gamma}
    \sum_{i=1}^{N} \sum_{i \neq j}^{N} \, \left(\frac{\vartheta_{ij}}{\vartheta_0}\right)^{1 - \gamma}\!\!(\Delta \Omega)^2,
\end{equation}
where $\Delta \Omega$ denotes the solid angle of the small pixels.
This definition ensures that $C_1 = 1$. We evaluate 
$C_\gamma$ at the large cell of $N_{side} = 16$ and split it into three different smaller cell sizes of $N_{side} = 1024, 2048, 4096$ corresponding to cell sizes of $0.57$, $0.28$, $0.14$ $deg^2$, respectively. The result of $C_\gamma$ as function of $\gamma$ is shown in Fig.~\ref{fig:Cgamma}.
It shows that for $\gamma$ up to $\sim$2, $C_{\gamma}$ converges. With the higher values of $\gamma$ we need to split the cells to smaller solid angles, in order to converge. Since at almost all values of $\gamma \leq 2$, after applying SNR cuts, the $C_{\gamma}$ values converge, we do not need to split the cells to smaller solid angles.

\end{document}